\documentclass[a4paper,11pt]{article}
\pdfoutput=1 

\usepackage{jheppub} 

\usepackage[T1]{fontenc} 

\usepackage{amsmath,amsfonts,amsbsy,amssymb,array,accents,dsfont}
\usepackage{enumerate,latexsym,graphicx}
\usepackage{xcolor,tikz}
\usepackage{tcolorbox}
\usepackage{mathtools}

\usepackage{subfigure}
\usepackage{stmaryrd}

\def\cV{\mathcal {V}} \def\cW{\mathcal {W}}

\newcommand{\beq}{\begin{equation}}
\newcommand{\eeq}{\end{equation}}
\newcommand{\bea}{\begin{eqnarray}}
\newcommand{\eea}{\end{eqnarray}}

\newcommand{\vep}{\varepsilon}
\newcommand{\ep}{\epsilon}

\newcommand{\der}{\partial}

\newcommand{\nn}{\nonumber}

\newcommand{\N}{{\cal{N}}}

\newcommand{\Bw}{\boldsymbol{\w}}

\usepackage{braket}
\usepackage{tikz}
\usetikzlibrary{matrix,calc,positioning,decorations.markings,decorations.pathmorphing,decorations.pathreplacing}
\usetikzlibrary{arrows,cd}
\usepackage{bbding}
\usetikzlibrary{positioning}
\tikzset{>=stealth}

\newcommand{\psisl}{\psi}
\newcommand{\psisu}{\chi}

\newcommand{\del}{\partial}

\newcommand{\qqquad}{\;, \quad\qquad}  

\newcommand{\RR}{\mathbb{R}}
\newcommand{\ZZ}{\mathbb{Z}}

\newcommand{\Aa}{\mathcal{A}}
\newcommand{\Bb}{\mathcal{B}}
\newcommand{\Cc}{\mathcal{C}}

\newcommand{\Nn}{\mathcal{N}}
\newcommand{\Oo}{\mathcal{O}}

\newcommand{\Vv}{\mathcal{V}}

\newcommand{\Yy}{\mathcal{Y}}
\newcommand{\Ww}{\mathcal{W}}

\newcommand{\R}{\ensuremath{\mathbb{R}}}

\newcommand{\w}{\omega}

\usepackage{color}

\usepackage[normalem]{ulem}  

\definecolor{darkred}{rgb}{0.6,0,0}
\definecolor{darkblue}{rgb}{0,0,0.6}

\newcommand\p{\partial}

\newcommand{\be}{\begin{equation}}
\newcommand{\ee}{\end{equation}}

\usepackage[bbgreekl]{mathbbol}
\usepackage{amsfonts}
\DeclareSymbolFontAlphabet{\mathbb}{AMSb} 
\DeclareSymbolFontAlphabet{\mathbbl}{bbold} 

\title{\boldmath Spectral flow and the exact AdS$_3/$CFT$_2$ chiral ring  
}

\author[a
]{Sergio Iguri,}
\author[b]{Nicolas Kovensky}
\author[c]{and Juli\'an H.~Toro}

\affiliation[a]{Instituto de Astronomía y Física del Espacio (IAFE) - CONICET and Facultad de Ciencias Exactas y Naturales, Universidad de Buenos Aires, Ciudad Universitaria, 1428 Buenos Aires, Argentina.}
\affiliation[b]{Institut de Physique Th\'eorique, Universit\'e Paris Saclay, CEA, CNRS, Orme des Merisiers, 91191 Gif-sur-Yvette CEDEX, France.}
\affiliation[c]{Instituto de Investigaciones Matemáticas Luis A. Santaló, CONICET - Universidad de Buenos Aires, Ciudad Universitaria, 1428 Buenos Aires, Argentina.}
\emailAdd{siguri@iafe.uba.ar}
\emailAdd{nicolas.kovensky@ipht.fr}
\emailAdd{jtoro@dm.uba.ar}

\abstract{
We compute all worldsheet three-point functions involving spectrally-flowed operators in chiral multiplets of the space-time theory for strings in AdS$_3\times$S$^3\times$T$^4$,  thus completing the analysis of the full AdS$_3$/CFT$_2$ chiral ring.  
We make use of techniques recently developed for the bosonic sector, based on holomorphic covering maps from the worldsheet to the AdS$_3$ boundary. We highlight the role of the so-called series identifications when dealing with the complications originated by picture-changing spectrally-flowed states. We find an exact agreement with the predictions from 
the holographic CFT at the symmetric orbifold point, including both extremal and non-extremal correlators.  
}

\newcommand{\of}[1]{\left(#1\right)}
\newcommand{\off}[1]{\left[#1\right]}
\newcommand{\offf}[1]{\left\{#1\right\}}  
\newcommand{\T}[1]{\Tilde{#1}}
\begin{document} 
\maketitle
\flushbottom

\section{Introduction}
\label{sec: intro}

Exact descriptions are available on both sides of the AdS$_3$/CFT$_2$ correspondence. On the one hand, the holographic CFT is best  understood at the symmetric orbifold  point in moduli space. On the other, for pure Neveu-Schwarz (NS) fluxes the worldsheet theory is solvable, the main ingredient being the  Wess-Zumino-Witten (WZW) model based on the universal cover of SL(2,$\R$). However, for general values of $n_5$ --- the number of NS5-brane sources --- the latter is \textit{not} dual to a symmetric orbifold CFT\footnote{The exception corresponds to the $n_5=1$ case, for which the strings become tensionless and the holographic CFT reduces to Sym$^{n_1}\left(T^4\right)$, where $n_1$ is the number of F1-string sources  \cite{Giribet:2018ada,Gaberdiel:2018rqv,Eberhardt:2018ouy,Eberhardt:2020bgq}. BPS correlators in this context were studied in \cite{Dei:2019iym,Gaberdiel:2022oeu}. }, and therefore, when comparing these two theories, one must restrict to observables which are protected by non-renormalization theorems \cite{deBoer:2008ss,Baggio:2012rr}. The most relevant examples are the spectrum of chiral primary operators and their three-point functions. 

This program was initiated in \cite{Dabholkar:2007ey,Gaberdiel:2007vu}, where the authors determined some worldsheet three-point functions matching those predicted in  \cite{Jevicki:1998bm,Lunin:2000yv,Lunin:2001pw}. The computation was done only for states in the spectrally unflowed sector of the SL(2,$\R$)-WZW model, which captures the low-lying spacetime chiral primaries with weights $h \leq (n_5+1)/2$. Worldsheet vertex operators with larger values of $h$ have non-trivial spectral flow charges \cite{Argurio:2000tb} and require a different treatment. Spectral flow constitutes one of the distinctive aspects of the model, and it introduces important complications for studying the corresponding correlation functions \cite{Maldacena:2000hw,Maldacena:2001km}, especially in the supersymmetric context \cite{Giribet:2007wp,Cardona:2009hk,Iguri:2022pbp}.

Obtaining the full set of superstring three-point functions in AdS$_3\times$S$^3\times $T$^4$ (or K3), including those where spectral flow is not conserved, remains an open problem. In this paper, we complete the holographic matching (at all orders in $\alpha'$) for the full  AdS$_3$/CFT$_2$ chiral ring, building on \cite{Iguri:2022pbp}. For this, we focus on the short-string sector of the theory and make use of recently developed techniques for computing bosonic three-point functions involving vertex operators with arbitrary spectral flow charges \cite{Eberhardt:2019ywk,Dei:2021xgh,Iguri:2022eat,Bufalini:2022toj}. We extend these methods in order to compute descendant correlators appearing after the ghost picture-changing procedure, mandatory for supersymmetric correlators. We further obtain all relevant fermionic correlation functions. The analysis of non-protected correlators with long-string insertions and the corresponding matching with the proposal of \cite{Eberhardt:2021vsx} are left for future work. 

As the AdS$_3$/CFT$_2$ correspondence constitutes one of the few scenarios where we can study holography far away from the supergravity regime, our results provide an important contribution to understanding the precise mechanism at play behind this duality. Furthermore, the techniques we present in this work will also be instrumental when considering applications to black holes and some of their microstates \cite{Martinec:2017ztd,Martinec:2018nco,Martinec:2019wzw,Martinec:2020gkv,Bufalini:2021ndn,Martinec:2022okx}. %
The computation of the first string correlators for these more involved worldsheet models was carried out recently in \cite{Bufalini:2022wyp,Bufalini:2022wzu}. These results can also be connected to the study of holography beyond AdS, little string theory, and the so-called single-trace $T\bar{T}$ deformations of two-dimensional CFTs \cite{Kutasov:2001uf,Giveon:1999tq,Giveon:2017myj,Asrat:2017tzd,Chakraborty:2018vja,Chakraborty:2019mdf,Apolo:2019zai,Georgescu:2022iyx}.

The paper is organized as follows. In Section \ref{sec: superstrings} we introduce the worldsheet theory for superstrings in AdS$_3\times$S$^3\times $T$^4$ and provide the definition of all vertex operators corresponding to spacetime chiral primaries, relegating some details to Appendix \ref{sec: appA}. In Section \ref{sec:primarycorrs} we review the recent results on correlators of primary fields with non-trivial spectral flow charges    in the bosonic SL(2,$\R$)-WZW model \cite{Dei:2021xgh,Bufalini:2022toj}, and then extend these results to the SU(2) and fermionic sectors. The main results of this paper are obtained in Section 
\ref{sec: short strings}, where we compute all fusion rules and structure constants of the AdS$_3$/CFT$_2$ chiral ring from the worldsheet theory, and show that they agree with the predictions from the holographic CFT exactly for all $n_5>1$. Section    
\ref{sec: discussion} contains our concluding remarks and outlook.


\section{Brief review of superstrings in AdS$_3\times$S$^3\times $T$^4$}
\label{sec: superstrings}

We start by briefly reviewing the basic definitions for type IIB superstrings on  AdS$_3\times$S$^3\times $T$^4$. We focus on the holographic dictionary for spacetime chiral primaries and define the corresponding vertex operators in the NSNS and RR sectors of the worldsheet theory for $n_5>1$. More details can be found in   \cite{Dabholkar:2007ey,Giribet:2007wp,Iguri:2022pbp}.

\subsection{Basic definitions}

String propagation in AdS$_3\times$S$^3\times $T$^4$ with NSNS fluxes is characterized by the  supersymmetric WZW model based on  SL(2,$\R$)$\times$SU(2)$\times$U(1)$^4$. In this section we discuss this model, mostly following the notation of \cite{Iguri:2022pbp}.

The affine SL(2,$\R$) and SU(2) currents and fermions are denoted as $J^{a}$, $\psi^a$, $K^a$ and $\chi^a$, respectively, with $a=0,1,2$. They satisfy the following OPEs: 
\begin{alignat}{4}
&J^a(z)J^b(w) &&\sim \frac{\frac{n_5}{2}\eta^{ab}}{(z-w)^2}+\frac{i\epsilon^{ab}{}_{c}J^c(w)}{z-w},\quad&&K^a(z)K^b(w) &&\sim \frac{\frac{n_5}{2}\delta^{ab}}{(z-w)^2}+\frac{i\epsilon^{ab}{}_{c}K^c(w)}{z-w},\\
&J^a(z)\psi^b(w)&&\sim \frac{i\epsilon^{ab}{}_{c}\psi^c(w)}{z-w},\quad &&K^a(z)\chi^b(w)&&\sim \frac{i\epsilon^{ab}{}_{c}\chi^c(w)}{z-w},\\
&\psi^a(z)\psi^b(w)&&\sim \frac{\frac{n_5}{2}\eta^{ab}}{z-w},\quad &&\chi^a(z)\chi^b(w)&&\sim \frac{\frac{n_5}{2}\delta^{ab}}{z-w},
\end{alignat}
where $n_5$ is the level in both cases, while $\epsilon^{012}=1$, $\eta^{ab} = \eta_{ab} = (-++)$ and $\delta^{ab}=\delta_{ab} = (+++)$. As usual, we define the ladder operators as $J^\pm = J^1 \pm i J^2$, and similarly for $K^\pm$, $\psi^\pm$ and $\chi^\pm$. 
The supersymmetric currents split as 
\beq
J^a = j^a + \hat{\jmath}^a\qqquad
K^a = k^a + \hat{k}^a,
\eeq
where $j^a$ and $k^a$ generate bosonic affine algebras SL(2,$\R$)$_{k}$ and SU(2)$_{k'}$ with shifted levels $k = n_5 + 2$ and $k' = n_5-2$,  while 
\beq 
\hat{\jmath}^a = -\frac{i}{n_5}\epsilon^{a}{}_{bc}\psi^b \psi^c\qqquad 
\hat{k}^a = -\frac{i}{n_5}\epsilon^{a}{}_{bc}\chi^b \chi^c \,, \label{Fermionic Currents}
\eeq
generate fermionic SL(2,$\R$)$_{-2}$ and SU(2)$_{2}$ algebras, decoupled from the bosonic sector. The free bosons and fermions associated with the $T^4$ directions are written as $Y^i$ and $\lambda^i$, respectively, with $i=6,\dots,9$.
%
%
It becomes very convenient to bosonize the fermions by introducing bosonic fields $H_I$,  $I=1,\dots 5$, such that\footnote{To be precise, one has to refine this definition slightly in order to keep track of the cocycle factors. This amounts to replacing $H_I$ by $\hat{H}_I$ with 
$\hat{H}_I = H_I + \pi \sum_{J<I} N_J$, where $N_J \equiv \oint i\del H_J $  
so that 
$       e^{i a \hat{H}_I} e^{i b \hat{H}_J} = e^{i b \hat{H}_J} e^{i a \hat{H}_I} \, e^{i \pi a b}$ if $I > J$ \cite{Dabholkar:2007ey}. } 
\begin{subequations}
\begin{gather}
	 \psisl^\pm = \sqrt{n_5} \, e^{\pm iH_1} 
	 \, , \quad 
	\psisu^\pm = \sqrt{n_5} \, e^{\pm i H_2} 
	 \, , \quad
	 \lambda^{6} \pm i \lambda^7 = e^{\pm i H_4} 
	 \, , \quad 
	 \lambda^{8} \pm i \lambda^9 = e^{\pm i H_5} \, ,  \label{psiH1xiH2} \\
	 \qquad \psisl^0 = \frac{\sqrt{n_5}}{2} \, \left(e^{i H_3} - e^{-iH_3} \right)  \qqquad \psisu^0 =  \frac{\sqrt{n_5}}{2}  \, \left(e^{i H_3} + e^{-iH_3} \right) \, .
\end{gather}
\end{subequations}  
with $ H_I^\dagger = H_I$ for $I \ne 3 $ and $ H_3^\dagger = - H_3 $. 


The stress tensor $T$ and supercurrent $G$ characterizing the matter sector of the worldsheet CFT read 
\begin{eqnarray}
    T &=& \frac{1}{n_5} \left(j^a j_a - \psisl^a \der \psisl_a + 
    k^a k_a - \psisu^a \der \psisu_a 
    \right) + \frac{1}{2}
    \left(\der Y^i \der Y_i - \lambda^i \der \lambda_i\right),
    \label{TAdS3S3T4def}
    \\
    G &=& \frac{2}{n_5} \left(
    \psisl^a j_a + \frac{2i}{n_5}\psisl^0 \psisl^1 \psisl^2 + 
    \psisu^a k_a - \frac{2i}{n_5}\psisu^0 \psisu^1 \psisu^2
    \right) + i \:\lambda^i \der Y_i \, .
    \label{GAdS3S3T4def}
\end{eqnarray}
We also need to consider the standard $bc$ and $\beta \gamma$ ghost systems, leading to the BRST charge 
\begin{equation}
\label{eq:BRSToperator}
    {\cal{Q}} = \oint dz \left[ c \left(T + T_{\beta\gamma}\right) - \gamma \, G + c(\der c) b - \frac{1}{4} b \gamma^2\right] \, .
\end{equation}
The $\beta\gamma$ system is further bosonized as 
\begin{align}
	\beta = e^{-\varphi} \del \xi  \qqquad \gamma = \eta \:\! e^{\varphi} \,,
\end{align}
where $\varphi$ has a background charge $Q_\varphi = -2$, while $\xi(z)\eta(w) \sim (z-w)^{-1}$.
The spacetime supercharges can thus be written as 
\begin{equation}
    Q_\vep = \oint dz \, e^{-\varphi/2} S_\vep \qqquad   S_\vep = \exp \left(\frac{i}{2} 
    \sum_{I=1}^{5}\vep_I H_I\right),
    \label{supercharges}
\end{equation}
where $S_\vep$ are spin fields and $\vep_I=\pm 1$. BRST-invariance and the mutual locality constraints impose $\vep_1 \vep_2 \vep_3 = \vep_4 \vep_5 = 1$, leading to the 
supercharges of the spacetime $\Nn=(4,4)$ superconformal algebra. Finally, the zero modes of the worldsheet SU(2) currents are identified with those of the R-symmetry of the boundary theory. 

\subsection{Short string spectrum and  vertex operators: the NS sector}

We first focus on the NSNS sector of the theory. Vertex operators holographically dual to the low-lying chiral primaries of the holographic CFT are expressed as \cite{Gaberdiel:2007vu,Dabholkar:2007ey}
\begin{subequations}
\label{defVW0}
\bea
&\cV_{j}(x,u) &= e^{-\varphi} \psi(x) V_{j}(x)W_{j-1}(u),\\
&\cW_{j}(x,u) &= e^{-\varphi}V_{j}(x) \chi(u) W_{j-1}(u).
\eea 
\end{subequations}
in the canonical $(-1)$ ghost picture.  Given the weights $\Delta_j = -j(j-1)/n_5$ and $\Delta'_l = l (l+1)/n_5$, the Virasoro condition relates the SL$(2,\RR)_k$ and SU$(2)_{k'}$ spins $j$ and $l$, setting $l = j-1$. The bosonic primaries $V_j(x)$ and $W_l(u)$ satisfy 
\begin{alignat}{2}
&j^a(z) V_j(x,w) \sim \frac{D^a_{x,j}V_j(x,w)}{z-w}, \quad&&k^a(z) W_{l}(u,w) \sim  \frac{P^a_{u,l}W_{l}(u,w)}{z-w},
\end{alignat}
where
\begin{alignat}{6}
    &D^+_{x,j} &&= \p_x, \qquad &&D^0_{x,j} &&= x\p_x + j, \qquad &&D^-_{x,j} &&= x^2\p_x + 2jx \label{SL2diff0}\\
    &P^+_{u,l} &&= \p_u,\qquad &&P^0_{u,l} &&= u\p_u -l, \qquad &&P^-_{u,l} &&= -u^2\p_u +2lu.
\end{alignat}
Here we work in the so-called $x$-basis for SL(2,$\R$) and the isospin $u$-basis for SU(2).  Operators $W_{l}(u)$ have integer and half-integer spins in the range $0\leq l \leq k'/2$, while the corresponding modes $W_{l,n}$ have spin projections taking values $n=-l,-l+1, \dots,l-1,l$ \cite{Zamolodchikov:1986bd}. 
Operators $V_{j}(x)$ are associated to normalizable short string states for real spins satisfying $1/2 < j < (k-1)/2$, and, for the corresponding $\mathcal{D}_j^+$ ($\mathcal{D}_j^-$) representations, the modes $V_{j,m}$ have projections $m=j,j+1,\dots$ ($m=-j,-j-1,\dots$) \cite{Maldacena:2000hw}. 

The complex label $x$ is identified with the holomorphic coordinate on the AdS$_3$ boundary. Eqs.~\eqref{SL2diff0} show that in the bosonic theory $j_0^a$ realize the global sector of the spacetime Virasoro algebra, such that $j$ is identified with the spacetime weight $h$. In the supersymmetric model an analogous statement holds for the zero-modes $J_0^a$, so that $H =j$ for $\cW_{j}$, while for $\cV_{j}$, the fermions induce the shift $H = j-1$. 

Worldsheet operators associated to spacetime chiral primaries with $H>n_5/2$ belong to the so-called spectrally-flowed sectors of the theory and were constructed in \cite{Giribet:2007wp}. In terms of the mode algebra, spectral flow refers to the automorphisms
\beq
j^{0,\omega}_{n} = j^0_n - \frac{k}{2}\omega \delta_{n,0} 
\,, \quad
k^{0,\omega}_{n} = k^0_n + \frac{k'}{2}\omega \delta_{n,0} 
\,, \quad
j^{\pm,\omega}_{n} = j^\pm_{n\pm\omega}
\,, \quad
k^{\pm,\omega}_{n} = k^\pm_{n\pm\omega}\, ,
\eeq 
\beq
\psi^{0,\w}_n =\psi^0_n
\,, \qquad 
\chi^{0,\w}_n =\chi^0_n 
\,, \qquad \psi^{\pm,\omega}_n = \psi^\pm_{n\pm\omega}
\,, \qquad 
\chi^{\pm,\omega}_n = \chi^\pm_{n\pm\omega}. \nn
\eeq
with $\omega \in \ZZ$. Let us first focus on the SL(2,$\R$) sector. Due to the fact that AdS$_3$ is both Lorentzian and non-compact, spectral flow gives rise to additional physical states, i.e. it generates representations inequivalent to the unflowed ones considered above. The corresponding bosonic vertex operators are built upon the $m$-basis spectrally-flowed primaries $V^{\omega}_{j,m}$. The latter are affine primary fields with respect to the flowed currents $j^{a,\omega}$. Hence, for $\w>0$, we have 
\begin{subequations}
\bea
&j^{0}(z)V^{\omega}_{j,m}(w)&= (m+\frac{k}{2}\omega)\frac{V^{\omega}_{j,m}(w)}{z-w} + \cdots \, ,\\[1ex]
&j^{-}(z)V^{\omega}_{j,m}(w)&= (z-w)^{\omega-1} (m+j-1)V^{\omega}_{j,m-1}(w) + \cdots\, ,\label{J-Vw OPE}\\
&j^{+}(z)V^{\omega}_{j,m}(w)&= (m-j+1)\frac{V^{\omega}_{j,m+1}(w)}{(z-w)^{1+\omega}}+\sum_{l=0}^{\omega-1}\frac{\off{j^{0}_l V^{\omega}_{j,m}}(w)}{(z-w)^{l+1}} + \cdots\, , \label{J+Vw OPE}
\eea
\label{JVwOPE}
\end{subequations}
where the ellipses denote higher-order terms. As a consequence, these are the lowest-weight states in a (discrete) spin 
$h = 
m + k\omega/2$ 
representation of the zero-mode algebra, while the worldsheet weight becomes $
\Delta^\omega_j = -\frac{j(j-1)}{k-2} - \omega m - \frac{k }{4}\omega^2 $.
Importantly, $V^{\omega}_{j,m}$ and $V^{-\omega}_{j,-m}$ lead to the same spin, and they both contribute to the same $x$-basis operator, such that one can freely restrict to positive values of $\w$ in this picture. The precise definition can be written as  
\begin{equation}
    V_{j,h}^\w (x) = e^{x j_0^+}  
    V_{j, m}^\w e^{-x j_0^+}\,,
\end{equation}
which can be understood as translating the operator at the origin, namely $V_{j, m}^\w = V_{j, h }^\w (0)$. 
A similar story holds for the SL(2,$\R$) fermions. Fermionic spectrally-flowed $m$-basis primaries are nicely expressed in bosonized form as  
\beq
\psi^{-,\omega} = \sqrt{n_5}e^{-i(1+ \omega)H_1 }\qqquad\psi^{+,\omega} =\sqrt{n_5}e^{i(1- \omega)H_1 }\qqquad\psi^{0,\omega} = \psi^0 e^{-i\omega H_1}.
\label{flowedfermionsSL2}
\eeq
Note that,
$\psi^{0,\omega=1} = \frac{\sqrt{n_5}}{2}\hat{\jmath}^-$, hence $\psi^{0,\omega+1} = \frac{\sqrt{n_5}}{2}\hat{\jmath}^{-,\omega}$.
We define the $x$-basis spectrally-flowed fermion and fermionic current that will be useful below, 
\begin{equation}
    \psi^\w(x) = e^{x \hat{\jmath}_0^+}  
    \psi^{-,\w} e^{-x \hat{\jmath}_0^+} \ , \ \hat{\jmath}^\w(x) = e^{x \hat{\jmath}_0^+}  
    \hat{\jmath}^{-,\w} e^{-x \hat{\jmath}_0^+}. 
\end{equation}

Although including spectral flow in the SU(2) sector is, strictly speaking, unnecessary, for short string states it turns out to be useful. Hence, we also introduce the  operators $W_{l,l_\w}^\w (u)$, $\chi^\w (u)$, $\hat{k}^\w (u)$, which are completely analogous to their SL(2,$\R$) cousins. 

We now have all the relevant ingredients to write down the supersymmetric vertex operators corresponding to the spectrally-flowed versions of those in Eq.~\eqref{defVW0}, which read 
\begin{subequations}
\label{defVWw}
\bea
&&\cV^{\omega}_{j}(x,u) = \frac{1}{\sqrt{n_5}}e^{-\varphi} \psi^{\omega}(x) V^{\omega}_{j}(x) 
\chi^{\omega-1}(u) W^{\omega}_{j-1}(u) ,\\
&&\cW^{\omega}_{j}(x,u) = \frac{1}{\sqrt{n_5}}e^{-\varphi} \psi^{\omega-1}(x) V^{\omega}_{j}(x) 
\chi^{\omega}(u) W^{\omega}_{j-1}(u) .\label{Flowed Fields}
\eea 
\end{subequations}
Here we have used that, in the (discrete) flowed case, the BRST constraints force the bosonic primary operators to be lowest-weight \cite{Giribet:2007wp}, and further set $l=j-1$. We have also introduced the shorthands
\begin{equation}
V^{\omega}_{j}(x) \equiv V^{\omega}_{j,j_\w}(x) \,, \quad 
W^{\omega}_{l}(u) \equiv W^{\omega}_{l,l_\w}(u)
\label{defVjw(x)}
\end{equation}
with $j_\w = j+k \w/2$ and $l_\w = l+k' \w/2$. 
As a consequence, the spacetime weights  are given by 
\begin{equation}
    H \left[\cV^{\omega}_{j}\right] =  j-1+n_5 \w/2 \qqquad H \left[\cW^{\omega}_{j}\right] = 
    j+n_5 \w/2 \,.
    \label{hdefVWw}
\end{equation}

There are actually two additional families of supersymmetric flowed vertex operators one can construct, starting from highest-weight operators that we denote as $V^{\omega}_{-j}(x) \equiv V^{\omega}_{j,-j+k\w/2}(x)$ and $W^{\omega}_{-l}(x) \equiv W^{\omega}_{l,-l+k'\w/2}(x)$. Considering them separately will not be necessary.  
Indeed, the statement that representations in different spectral flow sectors are inequivalent only holds up to the so-called SL(2,$\R$) series identifications \cite{Maldacena:2000hw}. At the bosonic level, we have equivalence relations for highest/lowest-weight flowed primaries taking the following form:  
\beq
V^{\omega}_{j} (x) = \N(j)V^{\omega+1}_{-\of{\frac{k}{2}-j}} (x)\, \qqquad W^{\omega}_{l} (u)= W^{\omega+1}_{-\of{\frac{k'}{2}-l}} (u) \, ,
\label{seriesIdBosMbasis}
\eeq
where we have also included the SU(2) counterpart. The coefficient $\N(j)$ is defined in terms of the reflection coefficient appearing in the SL(2,$\R$) two-point function, see Appendix \ref{sec: appA}. 
Eq.~\eqref{seriesIdBosMbasis} implies that, in the supersymmetric theory, the extra families of operators alluded to above do not lead to additional states, so it is enough to work with those in \eqref{defVWw}. We refer the reader to \cite{Iguri:2022pbp} for more details.

Fermionic representations also satisfy a similar identification. The $\hat{\jmath} = -1$ flowed field $\psi^\w(x)$ is mapped to a $\hat{\jmath}=0$ flowed field that we denote by $\hat{\psi}^{\w}(x)$. More precisely, this reads
\begin{equation}
    \psi^{\w}(x) = \sqrt{n_5}\hat{\psi}^{\w+1}(x).
\end{equation}
The unflowed field $\hat{\psi}(x)$ is nothing but the fermionic identity.
This can be done similarly for the SU$(2)$ fermions.

The identities presented in Eq.~\eqref{seriesIdBosMbasis} played a prominent role in the computation of bosonic spectrally-flowed primary three-point functions  \cite{Bufalini:2022toj}, which we review in Sec.~\ref{sec:primarycorrs} below. As it turns out, the series identifications will also be a crucial ingredient in the present paper, as they will allow us to bypass the additional technical difficulties that arise when computing short-string three-point functions in the supersymmetric model involving vertex operators with non-trivial spectral flow charges \cite{Giribet:2007wp,Iguri:2022pbp}.

\subsection{The R sector}

Vertex operators in the Ramond sector of the theory are constructed by using the spin fields defined in Eq.~\eqref{supercharges}.
The AdS$_3\times$S$^3$ chirality is defined as $\vep = \vep_1\vep_2\vep_3$, and the  
GSO projection imposes $\vep_4\vep_5=\vep$. The relevant unflowed vertex operators then involve two SL(2,$\R$)$_{-2}\times$SU(2)$_2$ fields  
 of spins $(j,l)=(-1/2,1/2)$
, namely
\begin{equation}
    s_{\vep}(x,y) = e^{u\hat{k}^0_0 } e^{x\hat{\jmath}^0_0 } e^{\frac{i}{2} \left( -H_1-H_2+\vep H_3 \right) } e^{-x\hat{\jmath}^0_0 } e^{-u\hat{k}^0_0 } \qqquad \vep = \pm 1 \,.
    \label{Rxbasis}
\end{equation}
Cocycle factors are important for computing the RHS of \eqref{Rxbasis}.  More explicitly,  
BRST invariant unflowed states have spins $(j-1/2,l+1/2 = j-1/2)$ with respect to the supersymmetric currents $J^a$,  $K^a$, and take the form \cite{Kutasov:1998zh,Dabholkar:2007ey} 
\begin{equation}
    \Yy^{\ep}_{j}(x,u) = e^{-\frac{\varphi}{2}} s_{-}(x,u)V_j(x)W_{j-1}(u)e^{i\frac{\ep}{2}\of{H_4-H_5}}, \label{R vertex}
\end{equation}
where we have renamed $\vep_4 \to \ep$ for simplicity. 
The corresponding spectrally-flowed states are constructed as in the NS sector.  
One obtains 
\begin{equation}
    \Yy^{\ep,\w}_{j}(x,u) = e^{-\frac{\varphi}{2}}s^\w_{-}(x,u)V^\w_{j}(x)W^\w_{j-1}(u)e^{i\frac{\ep}{2}\of{H_4-H_5}} \label{R flowed vertex},
\end{equation}
where $s^\w_{-}(x,u)$ is built upon the $m$-basis flowed primary 
\begin{equation}
     s^{\w}_{---} = e^{-i\of{\frac{1}{2}+\w}H_1 -i\of{\frac{1}{2}+\w}H_2 -\frac{i}{2}H_3 }, 
\end{equation}
which is the extremal state in a spin ($-1/2-\w,1/2+\w$) representation of the fermionic zero-mode algebra. As a consequence, the spacetime weights are 
\begin{equation}
    H \left[\Yy^{\ep,\w}_{j}\right] =  j-1/2+n_5 \w/2 .
\end{equation}

\subsection{Holographic dictionary}

We now discuss the identification of each of these operators in terms of the boundary theory at the symmetric orbifold point, namely Sym$^{N}\left(T^4\right)$ with $N=n_1 n_5$, following \cite{Argurio:2000tb,Dabholkar:2007ey,Giribet:2007wp}. At large $N$, the holographic dictionary identifies single string states in the bulk with single cycle fields of the dual CFT. Hence, one needs to start from the twist fields, usually denoted $\sigma_n$, which must be dressed appropriately. In each twist sector, one finds four types of chiral primary operators, corresponding to which of the chiral operators of the seed theory are used in this dressing. In the $T^4$ case, these can be the identity operator, the two complex  fermions $\psi^{a}$ with $a=1,2$, and finally the product $\psi^1 \psi^2$. These are denoted as $O_n^-(x)$, $O_n^a(x)$ and $O_n^+(x)$, and their holomorphic weights are given by  
\begin{equation}
    H \left[O_n^{-} \right] = \frac{n-1}{2}  \,,\quad
    H \left[O_n^{a} \right] = \frac{n}{2} \,,\quad
    H \left[O_n^{+} \right] = \frac{n+1}{2} \,;\quad
 n=1,2,\dots.    \label{D1D5CFTweights3}
\end{equation}
These are local operators on the boundary, which are associated with $z$-integrated $x$-basis operators of the worldsheet theory. Moreover, one can combine the individual states with fixed R-charge in a given R-symmetry chiral multiplet by using the SU(2)$_{\rm R}$ currents, leading to the isospin variables introduced in \cite{Zamolodchikov:1986bd}. As the zero modes of the R-symmetry currents are identified with those of the worldsheet SU(2) currents, the isospin variable is identified with the coordinate $u$ used in the previous sections. Consequently, and up to the normalization, which will be discussed below, the holographic dictionary for the chiral primary sector is
\begin{equation}
    O_n^- (x,u) \leftrightarrow \cV^{\omega}_{j}(x,u) ,\quad 
    O_n^a (x,u) \leftrightarrow \Yy^{\ep,\w}_{j}(x,u),\quad 
    O_n^+ (x,u) \leftrightarrow \cW^{\omega}_{j}(x,u),  
\end{equation}
together with the identification 
\begin{equation}
 n \,=\, 2 j - 1 + n_5 \, \w \,. \label{njw}
\end{equation}
From the worldsheet point of view, the allowed ranges are 
\begin{equation}
    j = 1,\frac{3}{2},\dots,
\frac{n_5}{2}
,\qquad
\w = 0,1,\dots \, .
\label{jwrangecc}
\end{equation}
This shows that the worldsheet theory accounts for all chiral primaries of the holographic CFT, except for those  in the twisted sectors where $n$ is a (non-zero) multiple of $n_5$ \cite{Dabholkar:2007ey,Giribet:2007wp}. 
These would sit exactly at the lower boundary of the allowed range for $j$. However, at this point, the spectrum degenerates due to the presence of the zero-momentum states belonging to the continuous representations~\cite{Teschner:1997ft,Giveon:2001up}. 
The fact that the worldsheet theory for strings in AdS$_3$ fails to describe these states indicates that the NS5-F1 model sits at a singular point in moduli space \cite{Seiberg:1999xz}. This was shown to be resolved when RR fluxes are included, thus lifting the long string sector \cite{Eberhardt:2018vho}.

\subsection{Picture changing}
Given the ghost background charge, for three-point functions, it is necessary to compute the ghost picture $(0)$ version of the NSNS operators defined in the previous sections. Given a ghost picture $(-1)$ operator $\Oo^{(-1)}(z)$, we have 
\begin{equation}
    \Oo^{(0)}(z) = \lim_{w\rightarrow z}\of{e^{\varphi(w)} G(w)} \Oo^{(-1)}(z)\,.
\end{equation}
Then,
\begin{equation}
 \Vv^{\w,(0)}_j(x,u) = \Aa^{\w,1}_j + (-1)^\w \Aa^{\w,2}_j \,,
    \label{Vpicture0}
\end{equation}
with 
\begin{eqnarray}
   \Aa^{\w,1}_j &=& \off{j^{-}_{-1-\w}(x)- H\hat{\jmath}^{-}_{-1-\w}(x)} \hat{\psi}^\w(x) V^\w_j(x) \hat{\chi}^{\w}(u) W^\w_{j-1}(u)\,, \\[1ex]
   \Aa^{\w,2}_j &=&-\frac{1}{n_5}\off{k^{+}_{\w}(u)-H\hat{k}^+_w(u)}\psi^{\w}(x)V^{\w}_j(x)\chi^{\w}(u)W^\w_{j-1}(u) \,,
\end{eqnarray}
and
\begin{equation}
    \Ww^{\w,(0)}_j(x,u) = \Bb^{\w,1}_j + (-1)^\w \Bb^{\w,2}_j \,,
        \label{Wpicture0}
\end{equation}
with 
\begin{eqnarray}
    \Bb^{\w,1}_j &=& \off{k^{-}_{-1-\w}(u)-H\hat{k}^{-}_{-1-\w}(u)} \hat{\psi}^\w(x) V^\w_{j}(x)\hat{\chi}^{\w} (u) W^\w_{j-1}(u)(x) \,, \\
   \Bb^{\w,2}_j &=& \frac{1}{n_5} \off{j^{+}_{\w}(x) + H\hat{\jmath}^{+}_\w(x)} \psi^{\w}(x)V^\w_j(x)\chi^\w(u)W^\w_{j-1}(u)  \,,
\end{eqnarray}
where in each equation $H$ corresponds to the spacetime weight of the corresponding vertex operator. These were derived in \cite{Giribet:2007wp} and reviewed in \cite{Iguri:2022pbp}, although  here we have rewritten them in a slightly more symmetric way. Note that although $\Aa^{\w,1}_j$ and $\Aa^{\w,2}_j$ have the same total fermion parity, they differ in one unit between the SL$(2,\RR)$ and SU$(2)$ fermion sector, therefore only one of them can contribute in a given three-point function. The same holds for $\Bb^{\w,1}_j$ and $\Bb^{\w,2}_j$.

In the R sector, we need to express the field \eqref{R flowed vertex} in the $(-3/2)$ ghost picture for computing two-point functions. They are given by
\beq
\Yy^{\ep,\w,(-\frac{3}{2})}_j(x,u) =-\frac{\sqrt{n_5}
}{2j-1+n_5\w} e^{-\frac{3}{2}\varphi} s^{\w}_{+}(x,u) V^{\w}_j(x)W^{\w}_{j-1}(u) e^{i\frac{\ep}{2}\of{H_4-H_5}}
\,, \label{R flowed vertex -3/2}
\eeq
where $s^\w_{+}(x,u)$ is the $x$- and $u$-basis version of the field 
\beq
s^\w_{--+} = e^{-i\of{\frac{1}{2}+\w}H_1-i\of{\frac{1}{2}+\w}H_2+\frac{i}{2}H_3} \,.
\eeq 
Indeed, one can check that \eqref{R flowed vertex -3/2} satisfies
\begin{equation}
     \Yy^{\ep,\w}_{j}(x,u;z) = \lim_{w\rightarrow z}\of{e^{\varphi}G}(w)\Yy^{\ep,\w,(-\frac{3}{2})}_j(x,u;w) \,.
\end{equation}

\section{Spectrally-flowed primary correlators}
\label{sec:primarycorrs}

In this section, we compute the primary correlators needed for the supersymmetric three-point functions. We first review the recent results obtained for bosonic SL(2,$\R$) correlators involving vertex operators  with arbitrary spectral flow charges \cite{Eberhardt:2019ywk,Dei:2021xgh,Iguri:2022eat,Bufalini:2022toj}. We then extend these results to the SU(2) sector of the theory and also provide the relevant correlators involving fermions and spin fields.   

\subsection{SL$(2,\RR)$ sector and the $y$-basis}

An integral formula for spectrally-flowed three-point functions in the bosonic SL(2,$\R$)-WZW model at level $k>3$ was conjectured in \cite{Dei:2021xgh} and proved recently in \cite{Bufalini:2022toj}\footnote{A similar conjecture for four-point functions was then put forward in \cite{Dei:2021yom} and further analyzed in \cite{Dei:2022pkr}. }. The intuition comes partly from the $k=3$ case, which corresponds to the background with a single NS5-brane in the supersymmetric context\footnote{Here the RNS formalism breaks down since the bosonic SU(2) level $k'=n_5-2$ becomes negative, and one needs to use the so-called hybrid formalism \cite{Berkovits:1999im}}. There, the fundamental strings become effectively tensionless, and the holographic CFT is identified with the symmetric orbifold model Sym$^{n_1}\left(T^4\right)$ \cite{Gaberdiel:2018rqv,Giribet:2018ada,Eberhardt:2018ouy,Eberhardt:2019ywk,Eberhardt:2020bgq}. In particular,  worldsheet spectral flow is mapped to the spacetime twist via AdS/CFT. 

Spectrally-flowed correlators with at least one unflowed insertion were discussed in \cite{Iguri:2022pbp}. Here  we restrict to cases with $\w_i \geq 1$ for $i=1,2,3$, and take $\w_3\geq \w_{1,2}$ for concreteness. Twisted correlators in the symmetric orbifold theory are non-zero only when one can construct a holomorphic covering map $\Gamma[\w_1,\w_2,\w_3](z) \equiv \Gamma(z)$ satisfying 
\begin{equation}
\Gamma(z) \sim x_i + a_i (z-z_i)^{\w_i} + \cdots\qquad \text{when} \quad 
z\sim z_i\,,
\label{coveringmapexp}
\end{equation}
where the ellipsis indicates higher order terms in $(z-z_i)$. Holographically, the covering space is identified with the string worldsheet itself. For three-point functions, this can always be done when \cite{Lunin:2000yv} 
\begin{equation}
    \w_1+\w_2+\w_3 \in 2\mathbb{Z}+1 
    \qqquad \w_1+\w_2 > \w_3 - 1 \, . \label{conditionsmapwi}
\end{equation}
At large $n_1$ the main contribution comes from covering surfaces of genus zero. For the above three-point functions, such a map is unique. After fixing $(z_1,z_2,z_3)=(x_1,x_2,x_3)=(0,1,\infty)$ as usual, the coefficients $a_i$ appearing in Eq.~\eqref{coveringmapexp} take the following combinatorial form 
\begin{equation}
    a_i =  
    \left(
    \begin{array}{c}
        \frac{\w_i+\w_{i+1}+\w_{i+2}-1}{2}  \\
        \frac{-\w_{i}+\w_{i+1}+\w_{i+2}-1}{2}
    \end{array}\right)
   \left(
    \begin{array}{c}
        \frac{-\w_{i}+\w_{i+1}-\w_{i+2}-1}{2} \\
        \frac{\w_{i}+\w_{i+1}-\w_{i+2}-1}{2}
    \end{array}\right)^{-1}\,,
    \label{coveringmapcoeffs}
\end{equation}
where the subscripts are understood to be mod 3. The use of the map $\Gamma(z)$ combined with the \textit{local} Ward identities derived from the OPEs \eqref{JVwOPE} allowed the authors of \cite{Eberhardt:2019ywk} to derive the corresponding three-point functions in the $n_5=1$ worldsheet theory. 

As mentioned in the introduction, for $k>3$ ($n_5>1$) the holographic CFT does \textit{not} have a symmetric orbifold structure. Nevertheless, the same local Ward identities imply a set of recursion relations for $x$-basis correlators \cite{Eberhardt:2019ywk}. These become differential equations in the $y$ variable introduced in \cite{Dei:2021xgh}. For a given spin $j$ and spectral flow charge $\w$, the so-called $y$-basis operators $\T{V}_{j}^{\w}(x,y,z)$ are constructed by summing all $V_{jh}^{\w}(x,z)$ over the allowed values of $h$. The inverse relationship can be written as a Mellin-type transform, namely 
\begin{equation}
     V_{j,h}^{\w}(x,z) = 
     \int d^2y \,  y^{j-m-1}  \bar{y}^{j-\bar{m}-1} 
     \T{V}_j^\w (x,y,z)\, ,
     \label{Mellinytoxbasis}
\end{equation}  
which mirrors the relation between the $m$-basis and the $x$-basis for unflowed operators \cite{Maldacena:2001km}. 
For operators $\T{V}_{j}^\w(x,y,z)$, Eq.~\eqref{Mellinytoxbasis} shows that the modes $j^\pm_{\pm \w}$ and $j_0^0$ act as differential operators in the $y$ variable, namely
\begin{equation}
    j^+_\w \sim D^+_{y,j} \qqquad 
    j^0_0 \sim D^0_{y,j} + \frac{k}{2}\w \qqquad 
    j^-_{-\w} \sim D^-_{y,j}\,, \label{DiffOpsVwY}
\end{equation}
where we have used the notation of Eq.~\eqref{SL2diff0}. When \eqref{conditionsmapwi} is satisfied, the differential equations the $y$ implied by the local Ward identities are solved by the following expression: 
\begin{align}
\langle \T{V}_{j_1}^{\w_1}(y_1)
    \T{V}_{j_2}^{\w_2}(y_2)
    \T{V}_{j_3}^{\w_3}(y_3) \rangle & = N_{\rm odd}\, (y_1-a_1)^{-2j_1}
    (y_2-a_2)^{-2j_2}
    (y_3-a_3)^{-2j_3} \label{oddfinal2} \\
    &\quad \times \left(
    \w_1\frac{y_1+a_1}{y_1-a_1}
    +\w_2\frac{y_2+a_2}{y_2-a_2}
    +\w_3\frac{y_3+a_3}{y_3-a_3} -1
    \right)^{\frac{k}{2}-j_1-j_2-j_3}. \nn 
\end{align}  
Here and from now on, we omit the explicit dependence in $x_i$ and $z_i$ unless necessary. The constant $N_{\rm odd}$, which fixes the normalization, was derived in \cite{Iguri:2022eat}, and will be given below. Eq.~\eqref{oddfinal2} is written in terms of the so-called $y$-basis operators, defined as follows. 
The more familiar $x$-basis three-point functions can then be extracted by means of \eqref{Mellinytoxbasis}.
For discrete states this can be computed in terms of holomorphic and anti-holomorphic contour integrals, while for complex $j$ one needs to integrate over the full complex plane.  

Moreover, short strings are an important part of the spectrum, and, as discussed above, the series identifications in Eq.~\eqref{seriesIdBosMbasis} show that, for such states, $\w$ is not defined uniquely. As a consequence, correlators can be non-trivial even when no associated covering map exits. This is consistent with the fact that the holographic CFT is not expected to be an exact symmetric orbifold \cite{Eberhardt:2021vsx}. One can use the adjacent covering maps of \cite{Bufalini:2022toj} to compute even parity correlators, namely those with  
\begin{equation}
    \w_1 + \w_2 + \w_3  \, \in 2  \mathbb{Z}\qqquad
    \w_1+\w_2 > \w_3  \, .
    \label{wiEvenCases}
\end{equation}
This leads to 
\begin{eqnarray}
\label{eq: general even solution}
    \langle \T{V}_{j_1}^{\w_1}(y_1)
    \T{V}_{j_2}^{\w_2}(y_2)
    \T{V}_{j_3}^{\w_3}(y_3) \rangle &= & N_{\rm even}
    \left(
    1-\frac{y_2}{ a_2[\Gamma_3^+]} -\frac{y_3}{a_3[\Gamma_2^+]}+
    \frac{y_2 y_3}{a_2[\Gamma_3^-] a_3[\Gamma_2^+] }\right)^{j_1-j_2-j_3} \nn \\
    & \times &   \left(
    1-\frac{y_1}{a_1[\Gamma_3^+]}-\frac{y_3}{a_3[\Gamma_1^+]} + \frac{y_1 y_3}{ a_1[\Gamma_3^-]  a_3[\Gamma_1^+]}
    \right)^{j_2-j_3-j_1} \\
    &\times &  \left(
    1-\frac{y_1}{a_1[\Gamma_2^+]} -\frac{y_2}{a_2[\Gamma_1^+]} +\frac{y_1 y_2}{a_1[\Gamma_2^+]a_2[\Gamma_1^-]} \right)^{j_3-j_1-j_2}\,, \nn
\end{eqnarray}
where $a_i[\Gamma_j^+]$ denotes the coefficient $a_i$ of the covering map in which $\w_j$ is shifted upwards by one unit, while $N_{\rm even}$ is discussed below. 

As it turns out, one can show that Eqs.~\eqref{oddfinal2} and \eqref{eq: general even solution} are also valid for the so-called edge cases,  
\begin{eqnarray}
\label{wiedgecases}
    \w_3 = \w_1+\w_2 \quad \text{or} \quad  \w_3 = \w_1+\w_2 +1\, ,  
\end{eqnarray}
which correspond to correlators which have a well-behaved $m$-basis limit \cite{Bufalini:2022toj,Cagnacci:2013ufa}. Roughly speaking, this also holds for correlators with unflowed insertions, where some of the $y$-variables are absent. Hence, these results provide an integral expression for all non-zero correlators satisfying the fusion rules derived in \cite{Maldacena:2001km}, 
\begin{equation}
    \w_1+\w_2 \geq \w_3-1, 
    \label{fusionrules}
\end{equation}
where, again, we have $\w_3\geq\w_{1,2}$. As discussed in \cite{Dei:2021xgh}, these selection rules are encoded in the normalization factors $N_{\rm odd}$ and $N_{\rm even}$ in Eqs.~\eqref{oddfinal2} and \eqref{eq: general even solution}, which are defined in terms of the unflowed three-point functions $C(j_1,j_2,j_3)$  \cite{Teschner:1999ug,Maldacena:2001km} combined with certain combinatorial factors related to spectral flow \cite{Dei:2021xgh,Iguri:2022eat}. Importantly, they factorize as follows 
\begin{equation}
    N_{\rm even} = C\left(j_1,j_2,j_3\right) \tilde{N}_{\rm even}(j_i,\w_i)\,, \quad 
    N_{\rm odd} = \N(j_1) C\left(k/2-j_1,j_2,j_3\right) \tilde{N}_{\rm odd}(j_i,\w_i)\,.
\end{equation} 
We have collected the definitions of these objects as well as some of their properties in Appendix \ref{sec: appA}. 

\subsection{SU$(2)$ and fermionic sectors}

In the SU(2) sector, spectrally-flowed correlators are merely complicated linear combinations of primary and descendant unflowed correlators. However, we can use techniques analogous to those of \cite{Eberhardt:2019ywk,Dei:2021xgh,Bufalini:2022toj}
in order to compute them directly. Indeed, using the same covering maps, we find that three-point functions of flowed SU(2) primaries satisfy the same recursion relations as those of the SL(2,$\R$) model with the replacements $k \to -k'$ and $j_i \to -l_i$. Hence, we conclude that, after fixing the overall dependence in the worldsheet and isospin variables by means of the global Ward identities,  we have 
\begin{align}
\langle \T{W}_{l_1}^{\w_1}(v_1)
    \T{W}_{l_2}^{\w_2}(v_2)
    \T{W}_{l_3}^{\w_3}(v_3) \rangle & = N_{\rm odd}'\, (v_1-a_1)^{2l_1}
    (v_2-a_2)^{2l_2}
    (v_3-a_3)^{2l_3} \label{oddfinal2su2} \\
    &\quad \times \left(
    \w_1\frac{v_1+a_1}{v_1-a_1}
    +\w_2\frac{v_2+a_2}{v_2-a_2}
    +\w_3\frac{v_3+a_3}{v_3-a_3} -1
    \right)^{-\frac{k'}{2}+l_1+l_2+l_3}, \nn 
\end{align}  
for odd parity while correlators, and
\begin{eqnarray}
\label{eq: general even solution su2}
    \langle \T{W}_{l_1}^{\w_1}(v_1)
    \T{W}_{l_2}^{\w_2}(v_2)
    \T{W}_{l_3}^{\w_3}(v_3) \rangle &= & N_{\rm even}'
    \left(
    1-\frac{v_2}{ a_2[\Gamma_3^+]} -\frac{v_3}{a_3[\Gamma_2^+]}+
    \frac{v_2 v_3}{a_2[\Gamma_3^-] a_3[\Gamma_2^+] }\right)^{-l_1+l_2+l_3} \nn \\
    & \times &   \left(
    1-\frac{v_1}{a_1[\Gamma_3^+]}-\frac{v_3}{a_3[\Gamma_1^+]} + \frac{v_1 v_3}{ a_1[\Gamma_3^-]  a_3[\Gamma_1^+]}
    \right)^{-l_2+l_3+l_1} \\
    &\times &  \left(
    1-\frac{v_1}{a_1[\Gamma_2^+]} -\frac{v_2}{a_2[\Gamma_1^+]} +\frac{v_1 v_2}{a_1[\Gamma_2^+]a_2[\Gamma_1^-]} \right)^{-l_3+l_1+l_2}\,, \nn
\end{eqnarray}
for even parity correlators. The normalizations are given by 
\begin{equation}
    N_{\rm even}'(l_i,\w_i) = C'\left(l_1,l_2,l_3\right) \tilde{N}_{\rm even}' \,, \quad 
    N_{\rm odd}'(l_i,\w_i) = C'\left(k'/2-l_1,l_2,l_3\right) \tilde{N}_{\rm odd}'\,,
\end{equation}
where $C'(l_1,l_2,l_3)$ are the unflowed SU(2) three-point functions, defined in Appendix \ref{sec: appA} together with the factors $\tilde{N}_{\rm odd}'$ and $\tilde{N}_{\rm even}'$. 
In the expressions above we have introduced the $v$-basis for SU(2) operators, analogous to the $y$-basis used in the SL(2,$\R$) case, i.e.
\begin{equation}
    \T{W}_l(u,v,z) = \sum_{n,\bar{n} = -l}^l v^{l+n}\bar{v}^{l+\bar{n}} W_{l,n+\frac{k'}{2}\w,\bar{n}+\frac{k'}{2}\w}(u,z)\,.
\end{equation}


Correlators involving flowed fermions and spin fields are also necessary for computing three-point functions in the supersymmetric model. We note that $\psi^{\w}(x,z)$ ($\chi^\w(u,z)$) is the flowed version of a spin $\hat{\jmath}=-1$ ($\hat{l}=1$) unflowed fermion, and belongs to a spin $\hat{\jmath}_\w = -1-\w$ ($\hat{l}_\w = 1+\w$)  representation of the zero-mode algebra of SL$(2,\RR)_{-2}$ (respectively SU$(2)_{2}$). Similarly, the fermionic field $\hat{\psi}^\w(x,z)$ ($\hat{\chi}^\w(u,z)$) can be seen as the flowed version of the fermionic identity, with flowed spin $\hat{\jmath}_\w = -\w$ ($\hat{l}_\w = \w$). Finally, the spin fields $s_\pm^\w(x,u,z)$ belong to  SL(2,$\R$)$_{-2}\times$SU(2)$_2$ representations with flowed spins $(\hat{\jmath}_\w,\hat{l}_\w)=(-1/2-\w,1/2+\w)$, obtained by the application of spectral flow on a spin $(\hat{\jmath},\hat{l})=(-1/2,1/2)$ state. Consequently, all relevant spectrally-flowed fermionic correlators can be obtained from the bosonic formulas given above by inserting the corresponding spins and levels and taking care of the parity of the different fermion numbers. 

As a check, we can recover some of the expressions obtained in \cite{Giribet:2007wp} using free field techniques. We first compute the correlator with three $\hat{\psi}^\w(x,z)$ insertions, which is non-vanishing only for even $\w_1+\w_2+\w_3$. The quickest way to compute this is to take the \textit{even} parity case, namely  Eq.~\eqref{eq: general even solution},  and set $j_1=j_2=j_3=0$ and $k=-2$. The desired $x$-basis result is then obtained by computing the relevant residue, which can be done simply by setting $y_i=0$ for $i=1,2,3$, see Eq.~\eqref{Mellinytoxbasis}. We obtain 
\begin{equation}
    \braket{\hat{\psi}^{\w_1}\hat{\psi}^{\w_2}\hat{\psi}^{\w_3}} = P_{(\w_1+1,\w_2+1,\w_3+1)}^2, 
    \label{GastonF0}
\end{equation}
where we have fixed the worldsheet and boundary insertion points as usual. We can also work out the  slightly more complicated case
\begin{eqnarray}
    &\braket{\hat{\psi}^{\w_1}\psi^{0,\w_2}\psi^{0,\w_3}}  =\frac{1}{4}\p_{y_2}\p_{y_3}\braket{\hat{\tilde{\psi}}^{\w_1}(y_1)\tilde{\psi}^{\w_2}(y_2)\tilde{\psi}^{\w_3}(y_3)}|_{y_i=0} \nn \\
    & = P_{(\w_1+1,\w_2+1,\w_3+1)}P_{(\w_1+1,\w_2-1,\w_3-1)}-P_{(\w_1+1,\w_2-1,\w_3+1)}P_{(\w_1+1,\w_2+1,\w_3-1)},   
    \label{GastonF2}
\end{eqnarray}
where we identified
\begin{eqnarray}
     \psi^{0,\w}(x,z) = -\frac{1}{2}(j^{+}_\w \psi^{\w})(x,z) = -\frac{1}{2}\p_y \tilde{\psi}^\w(x,y,z)|_{y=0}.
\end{eqnarray}
Our results \eqref{GastonF0} and \eqref{GastonF2} precisely reproduce the functions $f^{(0)}$ and $f^{(2)}$ of \cite{Giribet:2007wp}\footnote{We note that our $y$-basis methods give slightly different results for the functions $f^{(1)}$ and $f^{(3)}$ of \cite{Giribet:2007wp}. The formulas we derive lead to the correct holographic matching, as will be shown in the next section.}.


\section{Chiral primary three-point functions and holographic matching}
\label{sec: short strings}

This section contains the main results of this paper. We compute all relevant short-string three-point functions with arbitrary spectral flow insertions and match our results with the predictions from the chiral ring of the holographic CFT at the symmetric orbifold point. For this, we first compute a family of descendant correlators, which appear as a result of the picture-changing procedure. These correlators, presented as the main technical obstacle in this context \cite{Giribet:2007wp,Iguri:2022pbp}, are obtained by the SL(2,$\R$) series identifications, in combination with the $y$-basis results reviewed above. 

\subsection{Spectrally-flowed correlators involving current insertions}

Let us first focus on the NS-NS-NS short-string three-point functions. The simplest way to compute these is by inserting two vertex operators with ghost picture (-1) and one with ghost picture (0). For states polarized in the AdS$_3$ directions, the corresponding operators were given in Eq.~\eqref{Vpicture0}. From their expression, one finds that it is not enough to know the SL(2,$\R$) primary correlators. Indeed, we must also compute correlators of the form 
\beq
\braket{V^{\w_1}_{j_1}(x_1,z_1)V^{\w_2}_{j_2}(x_2,z_2)\of{j^{\w_3}V^{\w_3}_{j_3}}(x_3,z_3)},
\label{currentcorrelators}
\eeq
where $\of{j^{\w}V^{\w}_{j}}(x,z)$ stands for 
\beq
\of{j^{\w}V^{\w}_{j}}(x,z) = e^{x j^+_0}\oint_z dw \frac{j^-(w) V^{\w}_{j,j}(z)}{(w-z)^{1+\w}}e^{-x j^+_0}.
\eeq
Here $V^{\w}_{j,j}(z)$ is a flowed primary $m$-basis operator, derived from an unflowed lowest-weight state, as opposed to an $x$-basis operator.  

Three-point functions such as those in Eq.~\eqref{currentcorrelators} are usually thought of as descendant correlators in the sense that they involve the action of the mode  $j^-_{-1-\w}$ on a vertex operator $V_{j}^{\w}(x,z)$, which is a negative mode of the current $j^-$ in the corresponding spectrally-flowed frame. However, by using the series identification \eqref{seriesIdBosMbasis} we can interpret a given (short string) built upon an unflowed lowest-weight state as an operator with an extra unit of spectral flow charge built upon an unflowed highest-weight state, including the usual spin replacement $j \to k/2-j$. Hence, we can write 
\beq
\of{j^{\w}V^{\w}_{j}}(x,z) = \N(j) e^{x j^+_0}\oint_z dw \frac{1}{(w-z)^{1+\w}}j^-(w) V^{\w+1}_{\frac{k}{2}-j,-(\frac{k}{2}-j)}(z)
e^{-x j^+_0}.
\eeq
Crucially, the current mode $j^-_{-1-\w}$ can be seen as a zero mode in the spectrally-flowed frame associated to an operator with charge $\w+1$. Inserting the OPE \eqref{J-Vw OPE} in the above expression leads to\footnote{Although we  mostly use the shorthand defined in Eq.~\eqref{defVjw(x)}, for vertex operators built from unflowed states with $m\neq j$ it is necessary to go back to the usual notation, i.e. $V_{j,h}^\w(x)$. } 
\beq
\of{j^{\w}V^{\w}_{j}}(x,z) = -\N(j) e^{x j^+_0}V^{\w+1}_{\tilde{\jmath},-\tilde{\jmath}-1}(z)
e^{-x j^+_0}=  -\N(j) V^{\w+1}_{\tilde{\jmath},\tilde{h}-1}(x,z),
\eeq
where 
\begin{equation}
\tilde{\jmath} = k/2 - j \,,\quad  \tilde{h}= -\tilde{\jmath} + k(\w+1)/2 \,.   
\end{equation} 
We can then express the relevant   correlation functions with current insertions  in terms of flowed primary correlators, namely 
\begin{equation}
\braket{V^{\w_1}_{j_1}V^{\w_2}_{j_2}\of{j^{\w_3}V^{\w_3}_{j_3}}} = -\N(j_3) 
\braket{V^{\w_1}_{j_1}V^{\w_2}_{j_2}V^{\w_3+1}_{\tilde{\jmath}_3,\tilde{h}_3-1}}\, . 
\label{Current Correlator}
\end{equation}
Note that we have not imposed any condition on the spectral flow charges involved in the correlator. We are only taking advantage of the constraints imposed by the Virasoro condition for spectrally-flowed $1/2$-BPS vertex operators, i.e. $m_i = j_i$. 
Even though the strategy at play could have been used before, its applicability would have been limited, as flowed primary three-point functions were only computed in full generality quite recently \cite{Dei:2021xgh,Bufalini:2022toj}.  

In order to calculate the RHS of \eqref{Current Correlator} we go back to the $y$-basis operators  used in the previous section.  Eq.~\eqref{DiffOpsVwY} implies that, after fixing the insertions as $(z_1,z_2,z_3)=(x_1,x_2,x_3)=(0,1,\infty)$, the relevant correlator can be expressed in terms of its $y$-basis cousin as follows: 
\beq
\braket{V^{\w_1}_{j_1}V^{\w_2}_{j_2}V^{\w_3+1}_{\tilde{\jmath}_3,\tilde{h}_3-1}} \hspace{-0.05cm}  = \hspace{-0.2cm} \lim_{y_3\rightarrow \infty} y_3^{k-2j_3}\braket{\T{V}^{\w_1}_{j_1}(y_1=0)\T{V}^{\w_2}_{j_2}(y_2=0)D^{-}_{y_3,\tilde{\jmath}_3}\T{V}^{\w_3+1}_{\tilde{\jmath}_3}(y_3)}.
\eeq
By using Eqs.~\eqref{oddfinal2} and \eqref{eq: general even solution} we find that
\begin{equation}
 \braket{V^{\w_1}_{j_1}V^{\w_2}_{j_2}V^{\w_3+1}_{\tilde{\jmath}_3,\tilde{h}_3-1}}  
    = \alpha_{\Bw} \lim_{y_3\rightarrow \infty} y_3^{k-2j_3}\braket{\T{V}^{\w_1}_{j_1}(y_1=0)\T{V}^{\w_2}_{j_2}(y_2=0)\T{V}^{\w_3+1}_{\tilde{\jmath}_3}(y_3)},
\end{equation}
where the coefficient is given by
\begin{align}
    \alpha_{\Bw} \equiv \begin{dcases}
        & \frac{2 a_3[\Gamma_{13}^{++}]\off{(\w_1-\w_2)(j_1-j_2)+(\w_3+1)(\frac{k}{2}-j_3)}}{\w_1+\w_3-\w_2+1} \,\, \text{if} \,\,
        \sum_{i=1}^3 \w_i \in 2\mathbb{Z}+1,\\
        & \frac{2a_3[\Gamma_3^+]\off{(1+\w_1+\w_2)j_3-(1+\w_3)(j_1+j_2)-\frac{k}{2}(\w_1+\w_2-\w_3)}}{\w_3-\w_2-\w_1} \,\, \text{if} \,\,
        \sum_{i=1}^3 \w_i \in 2\mathbb{Z}.
    \end{dcases}
\end{align}
Here $a_3[\Gamma^{++}_{13}]$ denotes the coefficient $a_3$ of the covering map $\Gamma[\w_1+1,\w_2,\w_3+1](z)$. Finally, and after using the series identification once more, we conclude that 
\beq
\braket{V^{\w_1}_{j_1}V^{\w_2}_{j_2}\of{j^{\w_3}V^{\w_3}_{j_3}}} = \alpha_{\Bw} 
\braket{V^{\w_1}_{j_1}V^{\w_2}_{j_2}V^{\w_3}_{j_3}}.\label{Current Correlator final}
\eeq
Recall that this result holds only for the discrete states considered in this paper. Analogous formulas hold for the SL(2,$\R$) fermionic sector, where we have to set $k=-2$, and also for the SU(2) bosons and fermions, for which the levels and spins appear with the opposite sign.

\subsection{NS-NS-NS three-point functions}

We now have all the necessary  ingredients for computing the NS-NS-NS string correlators with arbitrary spectral flow charges. This includes not only the extremal correlators (in the holographic CFT language), some of which were briefly discussed in \cite{Giribet:2007wp}, but also the non-extremal ones. 

For concreteness, we start with all three vertex operators polarized in the AdS$_3$ directions, namely $
\langle \Vv^{\w_1}_{j_1} \Vv^{\w_2}_{j_2} \Vv^{\w_3,(0)}_{j_3} \rangle$.  
We take the picture (0) operator to be the one with the largest spectral flow charge,  $\w_3\geq \w_{1,2}$, and further set $(x_1,x_2,x_3) = (u_1,u_2,u_3) = (z_1,z_2,z_3)  = (0,1,\infty)$. Additionally, we will also assume that  
\begin{equation}
    \w_3 < \w_1 + \w_2 + 1.
    \label{nonedge}
\end{equation}
The edge cases $\w_3 = \w_1 + \w_2 + 1$ will be discussed below. When \eqref{nonedge} holds, all $y$-basis (and $v$-basis) three-point functions involved in the supersymmetric computation turn out to be regular in the limits  $y_i\to 0$ (and $v_i \to 0$). We can therefore pick up the relevant poles in the corresponding transforms \eqref{Mellinytoxbasis} by setting $y_i = v_i = 0$ in all relevant $y$-basis and $v$-basis correlators. 

Let us go back to the correlator $\langle \Vv^{\w_1}_{j_1} \Vv^{\w_2}_{j_2} \Vv^{\w_3,(0)}_{j_3} \rangle$. The vertex operators involved in this three-point function were given in Eqs.~\eqref{defVWw} and \eqref{Vpicture0}. Fermion number counting shows that for even parity correlators, i.e.~those with $\sum_i \w_i \in 2\mathbb{Z}$, only the first term in \eqref{Vpicture0} gives a non-zero contribution, while only the second one is relevant when $\sum_i \w_i \in 2\mathbb{Z}+1$. In other words, in the former case, we have 
\beq
\langle \Vv^{\w_1}_{j_1} \Vv^{\w_2}_{j_2} \Vv^{\w_3,(0)}_{j_3} \rangle = \langle \Vv^{\w_1}_{j_1} \Vv^{\w_2}_{j_2} \Aa^{\w_3,1}_{j_3} \rangle \qqquad 
\sum_i \w_i \in 2\mathbb{Z}. 
\eeq
This factorizes into ghost, bosonic and fermionic correlators. The latter include a current insertion, giving  
\begin{equation}
    \langle \psi^{\w_1}\psi^{\w_2}\of{\hat{\jmath}^{\w_3}\hat{\psi}^{\w_3}}\rangle = - \frac{2a_3[\Gamma^+_3](2+\w_1+\w_2+\w_3)}{\w_3-\w_2-\w_1}\langle\psi^{\w_1}\psi^{\w_2}\hat{\psi}^{\w_3}\rangle, 
\end{equation}
where we have used \eqref{Current Correlator final} with $k \to \hat{k}= -2$, $j_{1,2} \to \hat{\jmath}_{1,2} = - 1$ and $j_3 \to \hat{\jmath}_3=0$. Hence, we get 
\begin{align}
    &\langle \Vv^{\w_1}_{j_1} \Vv^{\w_2}_{j_2} \Vv^{\w_3,(0)}_{j_3} \rangle = (h_1+h_2+h_3-2)
    \label{VVV0primary}\\
    &\quad \times\frac{2(1+\w_3)a_3[\Gamma_3^+]}{\w_1+\w_2-\w_3}\braket{V^{\w_1}_{j_1}V^{\w_2}_{j_2}V^{\w_3}_{j_3}}\langle\psi^{\w_1}\psi^{\w_2}\hat{\psi}^{\w_3}\rangle\braket{W^{\w_1}_{j_1-1}W^{\w_2}_{j_2-1}W^{\w_3}_{j_3-1}}\braket{\hat{\chi}^{\w_1}\hat{\chi}^{\w_2}\hat{\chi}^{\w_3}},\nn
\end{align}
where $h_i = j_i + n_5\w_i/2 = H_i+1$. The spectrally-flowed primary three-point functions in the different sectors can be evaluated explicitly by using the results of Section \ref{sec:primarycorrs}. Although the individual factors look somewhat complicated, the final result becomes extremely simple:
\begin{align}
    \langle \Vv^{\w_1}_{j_1} \Vv^{\w_2}_{j_2} \Vv^{\w_3,(0)}_{j_3} \rangle_{\text{even}} = (h_1+h_2+h_3-2)n_5 \Cc(j_i).
    \label{VVV0final}
\end{align}
where $\Cc(j_i)$ is the product of the SL$(2,\RR)_k$ and SU$(2)_{k'}$ \textit{unflowed} three-point functions.

Let us pause and briefly discuss this expression. It was shown in \cite{Dabholkar:2007ey} that for short strings the relation between the SL(2,$\R$) and SU(2) spins stemming from the Virasoro condition leads to important cancellations for the product of unflowed three-point functions contained in $\Cc(j_i)$, whose final expressions is given in Eq.~\eqref{defCproduct} above. Here we have found that such cancellations extend non-trivially to the spectrally-flowed sectors of the theory. Indeed, as a result of the structural similarities between spectrally-flowed correlators for  SL(2,$\R$) and SU(2), all  combinatorial factors coming from the different contributions in the second line of \eqref{VVV0primary} exactly cancel each other. The only dependence of the final expression \eqref{VVV0final} on the spectral flow charges $\w_i$ is contained in the overall prefactor. This is consistent with the partial results of \cite{Iguri:2022pbp}, which were obtained by very different methods.  

We now show that the result in Eq.~\eqref{VVV0final} holds also for odd parity correlators with $\sum_i \w_i \in 2\mathbb{Z}+1$. In these cases we have 
\beq
\langle \Vv^{\w_1}_{j_1} \Vv^{\w_2}_{j_2} \Vv^{\w_3,(0)}_{j_3} \rangle = \langle \Vv^{\w_1}_{j_1} \Vv^{\w_2}_{j_2} \Aa^{\w_3,2}_{j_3} \rangle,
\eeq
 By  using that 
\begin{align}
    \langle W^{\w_1}_{l_1}W^{\w_2}_{l_2}\of{k^{+}_{\w_3}W^{\w_3}_{l_3}}\rangle = & \, \langle \T{W}^{\w_1}_{l_1}(v_1=0)\T{W}^{\w_2}_{l_2}(v_2=0)\p_{v_3}\T{W}^{\w_3}_{l_3}(v_3)\Large{|}_{v_3=0}\rangle \\
    =& -\frac{2\off{\w_3(l_1+l_2-\frac{k'}{2})-(1+\w_1+\w_2)l_3}}{(1+\w_1+\w_2+\w_3)a_3}\braket{W^{\w_1}_{l_1}W^{\w_2}_{l_2}W^{\w_3}_{l_3}},\nn
\end{align}
where $a_3$ is from the map $\Gamma[\w_1,\w_2,\w_3]$, and
\begin{equation}
    \langle \hat{\chi}^{\w_1}\hat{\chi}^{\w_2}\of{\hat{k}^+_{\w_3}\chi^{\w_3}}\rangle=\frac{2}{a_3}\braket{\hat{\chi}^{\w_1}\hat{\chi}^{\w_2}\chi^{\w_3}},
\end{equation}
we find 
\begin{align}
     \langle \Vv^{\w_1}_{j_1} \Vv^{\w_2}_{j_2} \mathcal{A}^{\w_3,2}_{j_3}\rangle =& -\of{h_1+h_2+h_3-2}\\
     &\times 2\frac{\braket{V^{\w_1}_{j_1}V^{\w_2}_{j_2}V^{\w_3}_{j_3}}\braket{W^{\w_1}_{j_1-1}W^{\w_2}_{j_2-1}W^{\w_3}_{j_3-1}}\braket{\psi^{\w_1}\psi^{\w_2}\psi^{\w_3}}\braket{\hat{\chi}^{\w_1}\hat{\chi}^{\w_2}\chi^{\w_3}}}{(1+\w_1+\w_2+\w_3)a_3}\nn.
\end{align}
As in the even parity case, and after using the identity Eq.~\eqref{3pt Reflection},
inserting the explicit expressions for each factor leads to \begin{align}
    \langle \Vv^{\w_1}_{j_1} \Vv^{\w_2}_{j_2} \Vv^{\w_3,(0)}_{j_3} \rangle_{\text{odd}} = (h_1+h_2+h_3-2)n_5 \Cc(j_i).
    \label{VVV0final2}
\end{align} 

A similar computation can be carried out for NS-NS-NS correlators involving one, two or three states polarized in the  SU(2) directions. The final results for these cases will be given in Section \ref{sec:normcorrelators} below.

\subsubsection{Edge Cases}
The $y$-basis three-point functions \eqref{oddfinal2} and \eqref{eq: general even solution} are singular  in the limits $y_i \to a_i$ \cite{Eberhardt:2021vsx,Iguri:2022eat,Dei:2022pkr,Bufalini:2022toj}. For the edge case, i.e. when $\w_3 = \w_1 + \w_2 + 1$, we see from \eqref{coveringmapcoeffs} that  $a_3=0$. As a consequence, we must be careful when picking up the relevant residues at $y_3=0$, and use the Mellin-like transform \eqref{Mellinytoxbasis}, which for discrete states can be performed as a contour integral, namely
\begin{equation}
    V^{\w_3}_{j_3}(x_3,z_3) = \oint_0 \frac{dy_3}{y_3} \T{V}^{\w_3}_{j_3}(x_3,y_3,z_3),
\end{equation}
where, as is done throughout the paper, we have ignored the anti-holomorphic variables. Analogous formulas hold for the SU(2) and fermionic sectors.

The three-point functions we are interested in are given by
\begin{align}
    &\braket{\Vv^{\w_1}_{j_1}\Vv^{\w_2}_{j_2}\Vv^{\w_3,(0)}_{j_3}} = \braket{\Vv^{\w_1}_{j_1}\Vv^{\w_2}_{j_2}\Aa^{\w_3,2}_{j_3}}=\nn\\&=\braket{\Vv^{\w_1}_{j_1}\Vv^{\w_2}_{j_2}\offf{\frac{\psi^{\omega_3}(x)}{n_5}V^{\omega_3}_{j_3}(x) \left[  k^+_{\w_3} - (j_3-1+\frac{n_5\w_3}{2}) W^{\omega_3}_{j_3-1}\hat{k}^{+}_{\w_3} \right]}W^{\omega_3}_{j_3-1}\chi^{\omega_3}}.
    \label{edgecases2terms}
\end{align}
Looking at the first term, the bosonic SL$(2,\RR)_k$ and SU$(2)_{k'}$ correlators involved are expressed as 
\begin{equation}
    \langle V^{\w_1}_{j_1}V^{\w_2}_{j_2}V^{\w_3}_{j_3}\rangle =
    \N(j_1)C(k/2-j_1,j_2,j_3) \oint_0 dy_3 y_3^{-1+\alpha}\of{1+y_3\frac{(\w_3+1)!}{\w_1!\w_2!}}^{\beta}
\end{equation}
and 
\begin{align}
    \langle W^{\w_1}_{j_1-1}W^{\w_2}_{j_2-1}&\of{k^+_\w W^{\w_3}_{j_3-1}}\rangle = \\
    &= C'(k'/2-j_1+1,j_2-1,j_3-1) \oint_0 dv_3 v_3^{-3-\alpha}\of{1+v_3\frac{(\w_3+1)!}{\w_1!\w_2!}}^{-(1+\beta)} \nn
\end{align}
where we have identified $k^+_{\w_3} \rightarrow \p_{v_3}$ and defined $\alpha = j_1+j_2-j_3-\frac{k}{2}$ and $\beta = \frac{k}{2}-j_1-j_2-j_3$. Due to the SU$(2)$ selection rules contained in the structure constants (and reviewed in Appendix \ref{sec: appA}) $\alpha$ and $\beta$ must be integer numbers, hence the integrals involved in these bosonic correlators will be finite. On the other hand, the fermionic SU$(2)$ correlator gives
\begin{equation}
\braket{\hat{\chi}^{\w_1}\hat{\chi}^{\w_2}\chi^{\w_3}} = n_5\oint_0 dv_3 v_3 = 0. 
\end{equation}
The first term in the second line of \eqref{edgecases2terms} thus vanishes. 
The same happens with the second term after identifying $\hat{k}^+_{\w_3}$ as a $v_3$-derivative for the SU(2) fermionic three-point function. Therefore, we conclude that 
\begin{equation}
    \langle \Vv^{\w_1}_{j_1}\Vv^{\w_2}_{j_2}\Vv^{\w_1+\w_2+1,(0)}_{j_3}\rangle = 0.
\end{equation}
A similar procedure shows that 
\begin{equation}
   \langle \Ww^{\w_1}_{j_1}\Ww^{\w_2}_{j_2}\Vv^{\w_1+\w_2+1,(0)}_{j_3}\rangle = \langle \Ww^{\w_1}_{j_1}\Ww^{\w_2}_{j_2}\Ww^{\w_1+\w_2+1,(0)}_{j_3}\rangle = 0.
\end{equation}

Finally, we can show that the last possible edge correlator  vanishes as well, i.e. 
\begin{equation}
    \langle \Ww^{\w_1}_{j_1}\Vv^{\w_2}_{j_2}\Vv^{\w_1+\w_2+1,(0)}_{j_3}\rangle = 0 \label{WVV}.
\end{equation}
This involves a factor of the form
\begin{equation}
    \langle V^{\w_1}_{j_1}V^{\w_2}_{j_2}\of{j^{\w_3}V^{\w_3}_{j_3}}\rangle.
\end{equation}
For this specific choice of spectral flow charges, this can be computed by means of techniques similar to those used in  Sec. 3.3 of \cite{Bufalini:2022toj}. We have 
\begin{equation}
   \langle \oint_{\Cc}dz j(x_3,z) V^{\w_1}_{j_1}(x_1,z_1)V^{\w_2}_{j_2}(x_2,z_2)V^{\w_3}_{j_3}(x_3,z_3)\rangle \frac{(z-z_1)^{\w_1+1}(z-z_2)^{\w_2+1}}{(z-z_3)^{\w_3+1}} = 0,
\end{equation}
for any contour $\Cc$ that encircles all three insertion points $z_i$, as  the integrand is regular at infinity. 
Moreover, the OPEs \eqref{JVwOPE} show that the integrand is regular at $z=z_1$ and $z=z_2$, and only the residue $z_3$ contributes. Hence,  
\begin{equation}
    \langle V^{\w_1}_{j_1} V^{\w_2}_{j_2}\of{j^{\w_3}V^{\w_3}_{j_3}}\rangle|_{\w_3=\w_1+\w_2+1} = 0.
\end{equation}
Since the corresponding fermionic correlator vanishes analogously, Eq.~\eqref{WVV} holds.

\subsection{R-R-NS correlators}

We now move to the R-R-NS three-point functions. While the extremal cases were obtained in  \cite{Giribet:2007wp}, here we complete the analysis by computing all the non-extremal ones. These three-point functions 
 \begin{equation}
\braket{\Yy^{\ep_1,\w_1}_{j_1}\Yy^{\ep_2,\w_2}_{j_2}\Vv^{\w_3}_{j_3}} \qquad \text{and} \qquad \braket{\Yy^{\ep_1,\w_1}_{j_1}\Yy^{\ep_2,\w_2}_{j_2}\Ww^{\w_3}_{j_3}} \label{RRNS}
\end{equation}
are technically simpler since no picture changing is necessary. Hence, they can be obtained directly from the spectrally-flowed primary correlators of Section \ref{sec:primarycorrs}. 

We start with the non-edge cases as before, i.e. with $\w_3 < \w_1+\w_2 +1$. The only new pieces of information we need are the fermionic correlators involving spectrally-flowed spin fields, namely 
\begin{equation}
    \braket{s_{-}^{\w_1} s_{-}^{\w_2} \psi^{\w_3}\hat{\chi}^{\w_3}}\braket{e^{\frac{i\ep_1}{2}\of{H_4-H_5}}e^{\frac{i\ep_2}{2}\of{H_4-H_5}}}.
\end{equation}
The different sectors factorize up to an overall phase coming from the cocycle factors (see footnote 2 above). However, this can be ignored since there is a single contribution, and the phase will cancel out upon including the contributions from the anti-holomorphic sector.  
The torus correlators involving  $H_4$ and $H_5$ impose $\ep_1 = -\ep_2$. On the other hand, the SL$(2,\RR)$ and SU$(2)$ contributions give a product of a flowed three-point function with SL$(2,\RR)_{-2}$ spins $(\hat{\jmath}_1,\hat{\jmath}_2,\hat{\jmath}_3) = (-1/2,-1/2,-1)$ and SU$(2)_2$ spins $(\hat{l}_1,\hat{l}_2,\hat{l}_3)=(1/2,1/2,0)$. 
As for the NS-NS-NS cases, we find that all combinatorial factors related to spectral flow cancel out, leading to
\begin{equation}
\braket{\Yy^{\ep_1,\w_1}_{j_1}\Yy^{\ep_2,\w_2}_{j_2}\Vv^{\w_3}_{j_3}} = 
\braket{\Yy^{\ep_1,\w_1}_{j_1}\Yy^{\ep_2,\w_2}_{j_2}\Ww^{\w_3}_{j_3}}  = \sqrt{n_5}\, 
\Cc(j_i) \xi^{\ep_1,\ep_2}\,,
\end{equation}
where $ \xi$ is the Pauli matrix $\sigma_1$. Finally, we find that  R-R-NS three-point functions with $\w_3 = \w_1+\w_2+1$ vanish as in the previous section. 

\subsection{Normalized correlators}
\label{sec:normcorrelators}

As it was argued in \cite{Maldacena:2001km,Giribet:2007wp} and proved recently in \cite{Iguri:2022pbp}, in order to obtain a precise holographic matching, the NSNS vertex operators must be normalized as
\begin{equation}
    \mathbb{O}^{\omega}_{j}(x,\bar{x},u,\bar{u},z,\bar{z}) = \frac{\Oo^{\omega}_{j}(x,\bar{x},u,\bar{u},z,\bar{z})}{\sqrt{2c_\nu^{-1}n_5Q^2\of{2h-1}B(j)}},
    \label{ONormNS}
\end{equation}
where $h = j+n_5\w/2$, while $c_\nu$, $Q$ and $B(j)$ are given in Appendix \ref{sec: appA}. A similar computation shows that, due to the extra factor appearing in Eq.~\eqref{R flowed vertex -3/2}, for vertex operators in the RR-sector of the worldsheet theory we have
\begin{equation}
    \mathbb{Y}^{\ep,\bar{\ep},\w}_j(x,\bar{x},u,\bar{u},z,\bar{z}) = \sqrt{\frac{2h-1}{2c_\nu^{-1}n_5^2Q^2B(j)}} \Yy^{\ep,\bar{\ep},\w}_j(x,\bar{x},u,\bar{u},z,\bar{z}).
    \label{ONormR}
\end{equation}
In these expressions we have re-inserted the anti-holomorphic dependence. Consequently, and after including the usual overall constants for string three-point functions, the results derived in the previous sections imply that the full set of normalized (spacetime) chiral primary three-point functions take the following form: 
\begin{subequations}
\label{Final3ptspacetimeNS}
\bea
&\braket{\mathbb{V}^{\w_1}_{j_1}\mathbb{V}^{\w_2}_{j_3}\mathbb{V}^{\w_3,(0)}_{j_3}}&= \frac{1}{\sqrt{N}}\off{\frac{\of{h_1 + h_2 + h_3-2}^4}{(2h_1-1)(2h_2-1)(2h_3-1)}}^{1/2},\\
&\braket{\mathbb{W}^{\w_1}_{j_1}\mathbb{V}^{\w_2}_{j_2}\mathbb{V}^{\w_3,(0)}_{j_3}}&=  \frac{1}{\sqrt{N}}\off{\frac{\of{1+ h_1 - h_2 - h_3}^4}{(2h_1-1)(2h_2-1)(2h_3-1)}}^{1/2},\\
&\braket{\mathbb{W}^{\w_1}_{j_1}\mathbb{W}^{\w_2}_{j_2}\mathbb{V}^{\w_3,(0)}_{j_3}}&=  \frac{1}{\sqrt{N}}\off{\frac{\of{h_1+h_2-h_3}^4}{(2h_1-1)(2h_2-1)(2h_3-1)}}^{1/2},\\
&\braket{\mathbb{W}^{\w_1}_{j_1}\mathbb{W}^{\w_2}_{j_2}\mathbb{W}^{\w_3,(0)}_{j_3}}&=  \frac{1}{\sqrt{N}}\off{\frac{\of{h_1+h_2+h_3-1}^4}{(2h_1-1)(2h_2-1)(2h_3-1)}}^{1/2},
\eea
\end{subequations}
and
\begin{subequations}
\label{Final3ptspacetimeR}
\bea
&\braket{\mathbb{Y}^{\ep_1,\bar{\ep}_1,\w_1}_{j_1}\mathbb{Y}^{\ep_2,\bar{\ep}_2,\w_2}_{j_2}\mathbb{V}^{\w_3}_{j_3}}&=  \frac{1}{\sqrt{N}}\off{\frac{\of{2h_1-1}\of{2h_2-1}}{(2h_3-1)}}^{1/2}\xi^{\ep_1,\ep_2}\xi^{\bar{\ep}_1,\bar{\ep}_2},\\
&\braket{\mathbb{Y}^{\ep_1,\bar{\ep}_1,\w_1}_{j_1}\mathbb{Y}^{\ep_2,\bar{\ep}_2,\w_2}_{j_2}\mathbb{W}^{\w_3}_{j_3}}&=  \frac{1}{\sqrt{N}}\off{\frac{\of{2h_1-1}\of{2h_2-1}}{(2h_3-1)}}^{1/2}\xi^{\ep_1,\ep_2}\xi^{\bar{\ep}_1,\bar{\ep}_2},
\eea
\end{subequations}
where $h_i = j_i+n_5 \w_i/2$. The overall scaling with $N=n_1 n_5$ is obtained from \cite{Dabholkar:2007ey}  
\beq
\frac{1}{\sqrt{N}}= \frac{g_s}{\sqrt{v_4}} \sqrt{\frac{2\pi^5}{\nu \gamma(1+b^2)}}.
\eeq
Here $g_s$ is the string coupling, $v_4$ is the volume of $T^4$, $b^2 = n_5^{-1}$, and  $\nu$ can be seen as a free parameter of the WZW model, which was fixed holographically in \cite{Dabholkar:2007ey}\footnote{See however footnote 4 in \cite{Iguri:2022pbp} and the discussion in \cite{Eberhardt:2021vsx}.} and is given in the Appendix. 
Finally, the selection rules on the SU(2) spins $l_i$ and spectral flow charges $\w_i$ can be summarized as follows:
\beq
l_i \leq l_k + l_j  \quad \text{and} \quad \w_i \leq \w_k +\w_j 
\quad \forall \quad i,j,k=1,2,3.
\label{finalfusionrules}
\eeq

The fusion rules \eqref{finalfusionrules} and the final expressions for the structure constants of the AdS$_3/$CFT$_2$ chiral ring presented in Eqs.~\eqref{Final3ptspacetimeNS} and \eqref{Final3ptspacetimeR} are in exact agreement with all previous worldsheet results \cite{Dabholkar:2007ey,Gaberdiel:2007vu,Giribet:2007wp,Cardona:2009hk,Iguri:2022pbp}. Moreover, they precisely reproduce the holographic CFT computations at the symmetric orbifold point \cite{Jevicki:1998bm,Lunin:2000yv,Lunin:2001pw}. This concludes our analysis of short-string correlators with arbitrary spectral flow charges. 


\section{Discussion}
\label{sec: discussion}
In this work, we studied short-string three-point functions of type IIB superstrings in AdS$_3\times$S$^3\times$T$^4$. More precisely, we focused on the worldsheet description of the spacetime chiral ring in this instance of the AdS$_3$/CFT$_2$ correspondence, and when the number of NS5-brane sources is strictly larger than one. It was discussed previously in the literature \cite{Giribet:2007wp,Cardona:2009hk,Iguri:2022pbp} that important complications arise when computing the three-point functions involving vertex operator insertions with non-zero spectral flow charges in the RNS formalism. Here we have shown how to overcome these difficulties. 

Bosonic primary correlators with arbitrary spectral flow charges were obtained recently in \cite{Eberhardt:2019ywk,Dei:2021xgh,Iguri:2022eat,Bufalini:2022toj}. Nevertheless, supersymmetric correlators remained elusive, mainly due to the appearance of current insertions arising from the picture-changing procedure. These descendant correlators cannot be obtained by the usual techniques involving contour integrals due to the rather non-trivial OPEs, presented in Eq.~\eqref{JVwOPE}. 

Recently, in \cite{Iguri:2022pbp} all short-string $x$-basis correlators involving at least one unflowed vertex operator were computed by using an $m$-basis approach based on \cite{Cagnacci:2013ufa}. However, such a strategy turned out to be somewhat restrictive. 

In this paper, we computed all remaining short-string supersymmetric correlators. This was done by generalizing the $y$-basis techniques developed in \cite{Dei:2021xgh,Bufalini:2022toj}, in order to apply them not only to the relevant descendant correlators but also to the SU(2) and fermionic sectors. As was the case for \cite{Bufalini:2022toj}, the SL(2,$\R$) series identifications \eqref{seriesIdBosMbasis} are instrumental in our calculations.

The individual bosonic and fermionic  correlators give complicated expressions as a function of the spectral flow charges, which moreover depend on the parity of the total spectral flow. We showed that, in all cases, the supersymmetric computation conspires  so that almost all of these combinatorial factors disappear from the final result. This extends the analysis of \cite{Dabholkar:2007ey} to the spectrally-flowed sectors of the theory. The final expressions show no trace of the difference between even and odd total spectral flow appearing in the intermediate steps, as was expected from a boundary CFT perspective. 

The main results of this paper are the fusion rules and structure constants presented in Eqs.~\eqref{Final3ptspacetimeNS}, \eqref{Final3ptspacetimeR} and  \eqref{finalfusionrules}. They precisely agree with the predictions from the holographic CFT at the symmetric orbifold point \cite{Jevicki:1998bm,Lunin:2000yv,Lunin:2001pw}, and complete the exact  (large $N$) holographic matching of the full AdS$_3$/CFT$_2$ chiral ring. 

The techniques developed in this paper will be useful for studying more general supersymmetric correlators for strings in AdS$_3$, i.e.~those involving long string states.  Unfortunately, for such states, the Virasoro condition is less restrictive, and in order to obtain the $x$-basis three-point functions  it will be necessary to carry out the integrals appearing in the Mellin transform \eqref{Mellinytoxbasis} over the full complex plane \cite{Dei:2021xgh}. Long-string correlators are not protected by non-renormalization theorems. Consequently, and as opposed to the short string case, a holographic comparison only makes sense with the holographic CFT at the same point in the moduli space. An exact matching with the results derived from the holographic CFT put forward in \cite{Eberhardt:2021vsx} would provide conclusive evidence for this proposal. We leave this computation for future work. 

Our results are also important in the context of black holes in AdS$_3$ and the description of some of their microstates, as the correlators we have computed here constitute the main building blocks of those in the gauged WZW models constructed in \cite{Martinec:2017ztd,Martinec:2018nco,Martinec:2019wzw,Martinec:2020gkv,Bufalini:2021ndn,Martinec:2022okx,Bufalini:2022wyp,Bufalini:2022wzu}. These models are also related to the discussions of \cite{Kutasov:2001uf,Giveon:1999tq,Giveon:2017myj,Asrat:2017tzd,Chakraborty:2018vja,Chakraborty:2019mdf,Apolo:2019zai,Georgescu:2022iyx} concerning
little string theory, $T\bar{T}$ deformations, and holography beyond AdS.    

\acknowledgments

It is a pleasure to thank Davide Bufalini for discussions. The work of S.I.~and J.H.T.~is supported by CONICET.  
The work of N.K.~is supported by the ERC Consolidator Grant 772408-Stringlandscape.

\appendix

\section{Some additional definitions and properties}
\label{sec: appA}
Here we provide some definitions related to the SL(2,$\R$) and SU(2) WZW models. We first focus on the unflowed sectors. For the SU(2)$_{k'}$ model, the zero-mode representations are the highest-weight states $W_{l}(u,z)$, with integer or half-integer spins $0 \leq l \leq k'/2$ \cite{Zamolodchikov:1986bd}. For strings in AdS$_3$, one considers states $V_j(x,z)$ in the continuous and discrete principal series of SL(2,$\R$)$_{k}$, more precisely, those with spins $j\in 1/2+i\R$ and $1/2 < j < (k-1)/2$, respectively \cite{Maldacena:2000hw}. The corresponding two-point functions (omitting the anti-holomorphic dependence as in the bulk of the paper) are given by 
\begin{equation}
\braket{W_{l_1}(u_1,z_1)W_{l_2}(u_2,z_2)}= \delta_{l_{1},l_2}\frac{u_{12}^{2l_{1}}}{z_{12}^{2\Delta_1'}}\, ,    
\end{equation}
with 
\begin{equation}
    \Delta_{l}' = \frac{l(l+1)}{k'+2} \,,
\end{equation}
and 
\beq
\braket{V_{j_1}(x_1,z_1)V_{j_2}(x_2,z_2)}=\frac{1}{z_{12}^{2\Delta_{1}}}\off{\delta^2(x_1-x_2)\delta(j_1+j_2-1)+ B(j_1)\frac{\delta(j_1-j_2)}{x_{12}^{2j_1}}}\label{SL2 bosonic 2point} \, , 
\eeq
with 
\begin{equation}
    \Delta_{j} = -\frac{j(j-1)}{k-2} \,,
\end{equation}
and where 
\begin{equation}
    B(j)=-\frac{\nu^{1-2j}}{\pi b^2}\gamma(1-b^2(2j-1)) \, , \quad 
    \gamma(x) = \frac{\Gamma(x)}{\Gamma(1-\bar{x})}\, , \quad
    b^2 = (k-2)^{-1}
     \label{defBj} \, .
\end{equation} 
The parameter $\nu$ is fixed holographically as in \cite{Dabholkar:2007ey,Giribet:2007wp,Iguri:2022pbp}\footnote{See footnote 4 in \cite{Iguri:2022pbp}} as
\begin{equation}
    \nu = \frac{2\pi^5}{b^4\gamma(1+b^2)}.
\end{equation} Recall that $k=n_5+2$ and $k'=n_5-2$ in the supersymmetric context. The function $B(j)$ defines the coefficient $\N(j)$ appearing in the series identifications \eqref{seriesIdBosMbasis},
\begin{equation}
    \N(j) =\sqrt{\frac{B(j)}{B(k/2-j)}}.
\end{equation} 
For short-string states such as those considered in this paper, only the second term in the two-point function \eqref{SL2 bosonic 2point} contributes. 
Finally, the coefficient $c_\nu$ appearing in Eqs.~\eqref{ONormNS} and \eqref{ONormR}
takes the form 
\beq
c_\nu = \frac{\pi\gamma(1-b^2)}{\nu b^2}.
\eeq
As for the three-point functions, in the SU(2) case we have 
\begin{equation}
    \braket{W_{l_1}(u_1,z_1)W_{l_2}(u_2,z_2)W_{l_3}(u_3,z_3)} = C'(l_1,l_2,l_3) \frac{
    u_{12}^{l_1 + l_2 - l_3}
    u_{23}^{l_2 + l_3 - l_1}
    u_{13}^{l_1 + l_3 - l_2}
    }{
    z_{12}^{\Delta_1'+\Delta_2'-\Delta_3'}
    z_{23}^{\Delta_2'+\Delta_3'-\Delta_1'}
    z_{13}^{\Delta_1'+\Delta_3'-\Delta_2'}}\, ,
\end{equation}
with 
\begin{equation}
    C'(l_1,l_2,l_3) = \sqrt{\gamma(b'^2)} P(l+1) \prod_{i=1}^3 \frac{P(l-2l_i)}{P(2l_i)\sqrt{\gamma(b'^2(2l_i+1))}}\,,
\end{equation}
where $l=l_1+l_2+l_3$, $b'^2 = (k'+2)^{-1}$ and 
\begin{equation}
    P(l) = \prod_{n=1}^l \gamma(n b^2) \qqquad P(0) = 1.
\end{equation}
These expressions hold if $2l_i \leq l \leq k'$ and $l\in 2\mathbb{Z}$, otherwise the three-point function vanishes. 
In the SL(2,$\R$) case the three-point function is given by 
\begin{equation}
    \braket{V_{j_1}(x_1,z_1)V_{j_2}(x_2,z_2)V_{j_3}(x_3,z_3)} = C(j_1,j_2,j_3) \frac{
    x_{12}^{j_3-j_1 -j_2 }
    x_{23}^{j_1-j_2 -j_3 }
    x_{13}^{j_2-j_1 -j_3 }
    }{
    z_{12}^{\Delta_1'+\Delta_2'-\Delta_3'}
    z_{23}^{\Delta_2'+\Delta_3'-\Delta_1'}
    z_{13}^{\Delta_1'+\Delta_3'-\Delta_2'}}\,,
\end{equation}
with, following the conventions of \cite{Dabholkar:2007ey},
\begin{equation}
    C(j_1,j_2,j_3) = -\frac{b^{1+b^2}\Upsilon(b)}{2\pi^2\gamma(1+b^2)}\frac{\of{\nu b^{2b^2}}^{1-j}}{\Upsilon(b(j-1))}\prod_i^3 \frac{\Upsilon\of{b(2j_i-1)}}{\Upsilon\of{b(j-2j_i)}}
\end{equation}
where $j=j_1+j_2+j_3$,  and the upsilon function $\Upsilon(x)$ has an integral representation as\footnote{The relation between the upsilon function and the function $G$ used in \cite{Maldacena:2001km} can be found in \cite{Teschner:1999ug}.}
\begin{equation}
    \ln\of{\Upsilon(x)} = \int_0^\infty \frac{dt}{t}\off{\of{\frac{q}{2}-x}^2e^{-t}-\frac{\sinh^2\of{(\frac{q}{2}-x)\frac{t}{2}}}{\sinh\of{\frac{bt}{2}}\sinh\of{\frac{t}{2b}}}},
    \,, \quad 
    q= b +b^{-1}\,,
\end{equation}
if $0< {\rm Re}(x) < q$, and it is extended outside this range by means of the following properties: 
\begin{equation}
    \Upsilon(x+b) = b^{1-2bx}\gamma\of{bx}\Upsilon(x)\, \qqquad \Upsilon(x+\frac{1}{b}) = b^{-1+\frac{2x}{b}}\gamma\of{\frac{x}{b}}\Upsilon(x).
\end{equation}
%
As pointed out in  \cite{Dabholkar:2007ey},  a crucial simplification occurs for the product of SU(2) and SL(2,$\R$) structure constants appearing in (unflowed) supersymmetric short-string correlators. More explicitly, we have 
\beq
   \Cc(j_i) \equiv  C(j_i)C'(j_i-1) =
 \sqrt{\frac{b^2\gamma(-b^2)}{4\pi \nu }}\prod_{i=1}^3 \sqrt{B(j_i)} \equiv Q\prod_{i=1}^3 \sqrt{B(j_i)}.
\label{defCproduct}
\eeq
Furthermore, the product $\Cc(j_i)$ satisfies the identities
\begin{eqnarray}
    \N(j_1) \Cc(k/2-j_1,j_2,j_3) = \N(j_2) \Cc(j_1,k/2-j_2,j_3)= \N(j_3) \Cc(j_1,j_2,k/2-j_3).\label{3pt Reflection}
\end{eqnarray}
This ensures that the three-point functions in Eq.~\eqref{oddfinal2} satisfy have the correct exchange symmetry. 
\medskip

The spectrum of the SL(2,$\R$) model also contains states in spectrally-flowed representations. The two-point functions of the corresponding operators can be obtained by means of the so-called parafermion decomposition, giving 
\cite{Maldacena:2001km}
\begin{equation}
    \langle V_{j_1 h_1}^\w (x_1,z_1) V_{j_2 h_2}^\w (x_2,z_2) \rangle = \frac{\delta^2(h_1-h_2)}{z_{12}^{2\Delta_1}x_{12}^{2h_1}}\left[\delta (j_1+j_2-1) +  \frac{\pi \delta(j_1-j_2) B(j_1) \gamma(j_1+m_1)}{\gamma(2j_1) \gamma(1-j_1+m_1)}\right].
    \label{SL2 2point bosonic with flow}
\end{equation}
Recall that for a given vertex $V_{j h}^\w (x,z)$, $m$ is defined in terms of the spacetime weight and the spectral flow charge as $m=h-k\w/2$. As in the unflowed sector, the first term in \eqref{SL2 2point bosonic with flow} is irrelevant for our short-string vertex operators.  
Three-point functions of vertex operators with $\w>0$ are much more involved, and were derived only recently  \cite{Eberhardt:2019ywk,Dei:2021xgh,Iguri:2022eat,Bufalini:2022toj}. We have reviewed these results in Section \ref{sec:primarycorrs}, see Eqs.~\eqref{oddfinal2} and \eqref{eq: general even solution}. The corresponding normalizations include the factors 
\begin{equation}
    \tilde{N}_{\rm even}(j_i,\w_i) = P_{(\w_1,\w_2,\w_3)}^{j_1+j_2+j_3-k}P_{(\w_1+1,\w_2+1,\w_3)}^{j_3-j_2-j_1} P_{(\w_1,\w_2+1,\w_3+1)}^{j_1-j_2-j_3}P_{(\w_1+1,\w_2,\w_3+1)}^{j_2-j_3-j_1} 
\end{equation}
and
\begin{equation}
    \tilde{N}_{\rm odd}(j_i,\w_i) =   \of{\frac{P_{(\w_1-1,\w_2-1,\w_3-1)}}{\w_1+\w_2+\w_3-1}}^{\frac{k}{2}-j_1-j_2-j_3} 
    P_{(\w_1-1,\w_2,\w_3)}^{j_3+j_2-j_1-\frac{k}{2}} P_{(\w_1,\w_2-1,\w_3)}^{j_3-j_2+j_1-\frac{k}{2}}P_{(\w_1,\w_2,\w_3-1)}^{-j_3+j_2+j_1-\frac{k}{2}} \, ,
\end{equation}
where 
\be 
P_{\boldsymbol{\w}} = 0 \qquad \text{for} \qquad \sum_j \w_j < 2 \max_{i=1,2,3} \w_i \quad \text{or}\quad \sum_i \w_i \in 2\mathds{Z}+1\,,
\ee
with $\boldsymbol{\w} = (\w_1,\w_2,\w_3)$, otherwise  
\be
P_{\boldsymbol{\w}} =S_{\boldsymbol{\w}} \frac{G\left(\frac{-\w_1+\w_2+\w_3}{2} +1\right) G\left(\frac{\w_1-\w_2+\w_3}{2} +1\right) G\left(\frac{\w_1+\w_2-\w_3}{2} +1\right) G\left(\frac{\w_1+\w_2+\w_3}{2}+1\right)}{G(\w_1+1) G(\w_2+1) G(\w_3+1)}  \ , 
\label{Pw-definition}
\ee
where $G(n)$ is the Barnes G function
\be 
G(n)=\prod_{i=1}^{n-1} \Gamma(i)
\label{barnesG}
\ee
for positive integer values, while $S_{\boldsymbol{\w}}$ is a phase depending on $\boldsymbol{\w} \bmod 2$. For more details, see \cite{Dei:2021xgh}. 
We also need spectrally-flowed correlators in the SU(2) sector. Two-point functions read  
\begin{equation}
\braket{W_{l_1}^{\w_1}(u_1,z_1)W_{l_2}^{\w_2}(u_2,z_2)}= \delta_{l_{1},l_2}\delta_{\w_{1},\w_2}\frac{u_{12}^{2l_{1}}}{z_{12}^{2\Delta_1'}}\, ,   
\end{equation}
where we have restricted to the vertex operators relevant for this paper, i.e. those appearing in supersymmetric spectrally-flowed short-string states, built from unflowed lowest-weight states. 
The extension of the methods of \cite{Eberhardt:2019ywk,Dei:2021xgh,Iguri:2022eat,Bufalini:2022toj} for three-point functions to the SU(2) case was also given in Section \ref{sec:primarycorrs}, leading to Eqs.~\eqref{oddfinal2su2} and \eqref{eq: general even solution su2}. In that context, one obtains the normalization factors 
\begin{equation}
    \tilde{N}_{\rm even}'(l_i,\w_i) = P_{(\w_1,\w_2,\w_3)}^{-l_1-l_2-l_3+k'}P_{(\w_1+1,\w_2+1,\w_3)}^{-l_3+l_2+l_1} P_{(\w_1,\w_2+1,\w_3+1)}^{-l_1+l_2+l_3}P_{(\w_1+1,\w_2,\w_3+1)}^{-l_2+l_3+l_1} 
\end{equation}
and
\begin{equation}
    \tilde{N}_{\rm odd}'(l_i,\w_i) =   \of{\frac{P_{(\w_1-1,\w_2-1,\w_3-1)}}{\w_1+\w_2+\w_3-1}}^{-\frac{k'}{2}+l_1+l_2+l_3} 
    P_{(\w_1-1,\w_2,\w_3)}^{-l_3-l_2+l_1+\frac{k'}{2}} P_{(\w_1,\w_2-1,\w_3)}^{-l_3+l_2-l_1+\frac{k'}{2}}
    P_{(\w_1,\w_2,\w_3-1)}^{l_3-l_2-l_1+\frac{k'}{2}} \, .
\end{equation}

Supersymmetric short-string three-point functions are greatly simplified by the relation between SL(2,$\R$) and SU(2) spins $l_i = j_i-1$. Indeed, the product between the normalizations $\tilde{N}_{\text{odd}/\text{even}}$ and $\tilde{N}'_{\text{odd}/\text{even}}$ leads to several cancellations, and it even becomes spin-independent:
\begin{equation}
\tilde{N}_{\rm odd}(j_i,\w_i)\tilde{N}'_{\rm odd}(j_i-1,\w_i) \equiv \Nn_{\rm odd}=\left[\frac{P(\w_1-1,\w_2-1,\w_3-1)\prod_i^3P(\boldsymbol{\w}-\hat{e}_i)}{\w_1+\w_2+\w_3-1}\right]^{-1},
\end{equation}
\begin{equation}
    \tilde{N}_{\rm even}(j_i,\w_i)\tilde{N}'_{\rm even}(j_i-1,\w_i)  \equiv \Nn_{\rm even}= \left[ P(\w_1,\w_2,\w_3)\prod_{i<j}P(\boldsymbol{\w}+\hat{e}_i+\hat{e}_j)\right]^{-1}, 
\end{equation}
where $\hat{e}_1 = (1,0,0)$, $\hat{e}_2 = (0,1,0)$ and $\hat{e}_3 = (0,0,1)$. This mirrors the simplifications for the product of the SL$(2,\RR)$ and SU$(2)$ unflowed three-point functions highlighted in Eq.~\eqref{defCproduct} above. These identities are crucial for the analysis of Section \ref{sec: short strings}.

\bibliographystyle{JHEP}
\bibliography{refs}

\providecommand{\href}[2]{#2}\begingroup\raggedright\begin{thebibliography}{10}

\bibitem{Giribet:2018ada}
G.~Giribet, C.~Hull, M.~Kleban, M.~Porrati and E.~Rabinovici,
  \emph{{Superstrings on AdS$_{3}$ at $\mathcal{k} =$ 1}},
  \href{https://doi.org/10.1007/JHEP08(2018)204}{\emph{JHEP} {\bfseries 08}
  (2018) 204} [\href{https://arxiv.org/abs/1803.04420}{{\ttfamily
  1803.04420}}].

\bibitem{Gaberdiel:2018rqv}
M.~R. Gaberdiel and R.~Gopakumar, \emph{{Tensionless string spectra on
  AdS$_{3}$}}, \href{https://doi.org/10.1007/JHEP05(2018)085}{\emph{JHEP}
  {\bfseries 05} (2018) 085}
  [\href{https://arxiv.org/abs/1803.04423}{{\ttfamily 1803.04423}}].

\bibitem{Eberhardt:2018ouy}
L.~Eberhardt, M.~R. Gaberdiel and R.~Gopakumar, \emph{{The Worldsheet Dual of
  the Symmetric Product CFT}},
  \href{https://doi.org/10.1007/JHEP04(2019)103}{\emph{JHEP} {\bfseries 04}
  (2019) 103} [\href{https://arxiv.org/abs/1812.01007}{{\ttfamily
  1812.01007}}].

\bibitem{Eberhardt:2020bgq}
L.~Eberhardt, \emph{{Partition functions of the tensionless string}},
  \href{https://doi.org/10.1007/JHEP03(2021)176}{\emph{JHEP} {\bfseries 03}
  (2021) 176} [\href{https://arxiv.org/abs/2008.07533}{{\ttfamily
  2008.07533}}].

\bibitem{Dei:2019iym}
A.~Dei and L.~Eberhardt, \emph{{Correlators of the symmetric product
  orbifold}}, \href{https://doi.org/10.1007/JHEP01(2020)108}{\emph{JHEP}
  {\bfseries 01} (2020) 108}
  [\href{https://arxiv.org/abs/1911.08485}{{\ttfamily 1911.08485}}].

\bibitem{Gaberdiel:2022oeu}
M.~R. Gaberdiel and B.~Nairz, \emph{{BPS correlators for AdS$_{3}$/CFT$_{2}$}},
  \href{https://doi.org/10.1007/JHEP09(2022)244}{\emph{JHEP} {\bfseries 09}
  (2022) 244} [\href{https://arxiv.org/abs/2207.03956}{{\ttfamily
  2207.03956}}].

\bibitem{deBoer:2008ss}
J.~de~Boer, J.~Manschot, K.~Papadodimas and E.~Verlinde, \emph{{The Chiral ring
  of AdS(3)/CFT(2) and the attractor mechanism}},
  \href{https://doi.org/10.1088/1126-6708/2009/03/030}{\emph{JHEP} {\bfseries
  03} (2009) 030} [\href{https://arxiv.org/abs/0809.0507}{{\ttfamily
  0809.0507}}].

\bibitem{Baggio:2012rr}
M.~Baggio, J.~de~Boer and K.~Papadodimas, \emph{{A non-renormalization theorem
  for chiral primary 3-point functions}},
  \href{https://doi.org/10.1007/JHEP07(2012)137}{\emph{JHEP} {\bfseries 07}
  (2012) 137} [\href{https://arxiv.org/abs/1203.1036}{{\ttfamily 1203.1036}}].

\bibitem{Dabholkar:2007ey}
A.~Dabholkar and A.~Pakman, \emph{{Exact chiral ring of AdS(3) / CFT(2)}},
  \href{https://doi.org/10.4310/ATMP.2009.v13.n2.a2}{\emph{Adv. Theor. Math.
  Phys.} {\bfseries 13} (2009) 409}
  [\href{https://arxiv.org/abs/hep-th/0703022}{{\ttfamily hep-th/0703022}}].

\bibitem{Gaberdiel:2007vu}
M.~R. Gaberdiel and I.~Kirsch, \emph{{Worldsheet correlators in
  AdS(3)/CFT(2)}},
  \href{https://doi.org/10.1088/1126-6708/2007/04/050}{\emph{JHEP} {\bfseries
  04} (2007) 050} [\href{https://arxiv.org/abs/hep-th/0703001}{{\ttfamily
  hep-th/0703001}}].

\bibitem{Jevicki:1998bm}
A.~Jevicki, M.~Mihailescu and S.~Ramgoolam, \emph{{Gravity from CFT on S**N(X):
  Symmetries and interactions}},
  \href{https://doi.org/10.1016/S0550-3213(00)00147-4}{\emph{Nucl. Phys. B}
  {\bfseries 577} (2000) 47}
  [\href{https://arxiv.org/abs/hep-th/9907144}{{\ttfamily hep-th/9907144}}].

\bibitem{Lunin:2000yv}
O.~Lunin and S.~D. Mathur, \emph{{Correlation functions for M**N / S(N)
  orbifolds}}, \href{https://doi.org/10.1007/s002200100431}{\emph{Commun. Math.
  Phys.} {\bfseries 219} (2001) 399}
  [\href{https://arxiv.org/abs/hep-th/0006196}{{\ttfamily hep-th/0006196}}].

\bibitem{Lunin:2001pw}
O.~Lunin and S.~D. Mathur, \emph{{Three point functions for M(N) / S(N)
  orbifolds with N=4 supersymmetry}},
  \href{https://doi.org/10.1007/s002200200638}{\emph{Commun. Math. Phys.}
  {\bfseries 227} (2002) 385}
  [\href{https://arxiv.org/abs/hep-th/0103169}{{\ttfamily hep-th/0103169}}].

\bibitem{Argurio:2000tb}
R.~Argurio, A.~Giveon and A.~Shomer, \emph{{Superstrings on AdS(3) and
  symmetric products}},
  \href{https://doi.org/10.1088/1126-6708/2000/12/003}{\emph{JHEP} {\bfseries
  12} (2000) 003} [\href{https://arxiv.org/abs/hep-th/0009242}{{\ttfamily
  hep-th/0009242}}].

\bibitem{Maldacena:2000hw}
J.~M. Maldacena and H.~Ooguri, \emph{{Strings in AdS(3) and SL(2,R) WZW model
  1.: The Spectrum}}, \href{https://doi.org/10.1063/1.1377273}{\emph{J. Math.
  Phys.} {\bfseries 42} (2001) 2929}
  [\href{https://arxiv.org/abs/hep-th/0001053}{{\ttfamily hep-th/0001053}}].

\bibitem{Maldacena:2001km}
J.~M. Maldacena and H.~Ooguri, \emph{{Strings in AdS(3) and the SL(2,R) WZW
  model. Part 3. Correlation functions}},
  \href{https://doi.org/10.1103/PhysRevD.65.106006}{\emph{Phys. Rev.}
  {\bfseries D65} (2002) 106006}
  [\href{https://arxiv.org/abs/hep-th/0111180}{{\ttfamily hep-th/0111180}}].

\bibitem{Giribet:2007wp}
G.~Giribet, A.~Pakman and L.~Rastelli, \emph{{Spectral Flow in AdS(3)/CFT(2)}},
  \href{https://doi.org/10.1088/1126-6708/2008/06/013}{\emph{JHEP} {\bfseries
  06} (2008) 013} [\href{https://arxiv.org/abs/0712.3046}{{\ttfamily
  0712.3046}}].

\bibitem{Cardona:2009hk}
C.~A. Cardona and C.~A. Nunez, \emph{{Three-point functions in superstring
  theory on AdS(3) x S**3 x T**4}},
  \href{https://doi.org/10.1088/1126-6708/2009/06/009}{\emph{JHEP} {\bfseries
  06} (2009) 009} [\href{https://arxiv.org/abs/0903.2001}{{\ttfamily
  0903.2001}}].

\bibitem{Iguri:2022pbp}
S.~Iguri, N.~Kovensky and J.~H. Toro, \emph{{Spectral flow and string
  correlators in AdS$_3\times S^3 \times T^4$}},
  \href{https://doi.org/10.1007/JHEP01(2023)161}{\emph{JHEP} {\bfseries 2023}
  (2023) 161} [\href{https://arxiv.org/abs/2211.02521}{{\ttfamily
  2211.02521}}].

\bibitem{Eberhardt:2019ywk}
L.~Eberhardt, M.~R. Gaberdiel and R.~Gopakumar, \emph{{Deriving the
  AdS$_{3}$/CFT$_{2}$ correspondence}},
  \href{https://doi.org/10.1007/JHEP02(2020)136}{\emph{JHEP} {\bfseries 02}
  (2020) 136} [\href{https://arxiv.org/abs/1911.00378}{{\ttfamily
  1911.00378}}].

\bibitem{Dei:2021xgh}
A.~Dei and L.~Eberhardt, \emph{{String correlators on AdS$_{3}$: three-point
  functions}}, \href{https://doi.org/10.1007/JHEP08(2021)025}{\emph{JHEP}
  {\bfseries 08} (2021) 025}
  [\href{https://arxiv.org/abs/2105.12130}{{\ttfamily 2105.12130}}].

\bibitem{Iguri:2022eat}
S.~Iguri and N.~Kovensky, \emph{{On spectrally flowed local vertex operators in
  AdS$_3$}}, \href{https://doi.org/10.21468/SciPostPhys.13.5.115}{\emph{SciPost
  Phys.} {\bfseries 13} (2022) 115}
  [\href{https://arxiv.org/abs/2208.00978}{{\ttfamily 2208.00978}}].

\bibitem{Bufalini:2022toj}
D.~Bufalini, S.~Iguri and N.~Kovensky, \emph{{A proof for string three-point
  functions in AdS$_{3}$}},
  \href{https://doi.org/10.1007/JHEP02(2023)246}{\emph{JHEP} {\bfseries 02}
  (2023) 246} [\href{https://arxiv.org/abs/2212.05877}{{\ttfamily
  2212.05877}}].

\bibitem{Eberhardt:2021vsx}
L.~Eberhardt, \emph{{A perturbative CFT dual for pure NS\textendash{}NS
  AdS$_{3}$ strings}}, \href{https://doi.org/10.1088/1751-8121/ac47b2}{\emph{J.
  Phys. A} {\bfseries 55} (2022) 064001}
  [\href{https://arxiv.org/abs/2110.07535}{{\ttfamily 2110.07535}}].

\bibitem{Martinec:2017ztd}
E.~J. Martinec and S.~Massai, \emph{{String Theory of Supertubes}},
  \href{https://doi.org/10.1007/JHEP07(2018)163}{\emph{JHEP} {\bfseries 07}
  (2018) 163} [\href{https://arxiv.org/abs/1705.10844}{{\ttfamily
  1705.10844}}].

\bibitem{Martinec:2018nco}
E.~J. Martinec, S.~Massai and D.~Turton, \emph{{String dynamics in NS5-F1-P
  geometries}}, \href{https://doi.org/10.1007/JHEP09(2018)031}{\emph{JHEP}
  {\bfseries 09} (2018) 031}
  [\href{https://arxiv.org/abs/1803.08505}{{\ttfamily 1803.08505}}].

\bibitem{Martinec:2019wzw}
E.~J. Martinec, S.~Massai and D.~Turton, \emph{{Little Strings, Long Strings,
  and Fuzzballs}}, \href{https://doi.org/10.1007/JHEP11(2019)019}{\emph{JHEP}
  {\bfseries 11} (2019) 019}
  [\href{https://arxiv.org/abs/1906.11473}{{\ttfamily 1906.11473}}].

\bibitem{Martinec:2020gkv}
E.~J. Martinec, S.~Massai and D.~Turton, \emph{{Stringy Structure at the BPS
  Bound}}, \href{https://doi.org/10.1007/JHEP12(2020)135}{\emph{JHEP}
  {\bfseries 12} (2020) 135}
  [\href{https://arxiv.org/abs/2005.12344}{{\ttfamily 2005.12344}}].

\bibitem{Bufalini:2021ndn}
D.~Bufalini, S.~Iguri, N.~Kovensky and D.~Turton, \emph{{Black hole microstates
  from the worldsheet}},
  \href{https://doi.org/10.1007/JHEP08(2021)011}{\emph{JHEP} {\bfseries 08}
  (2021) 011} [\href{https://arxiv.org/abs/2105.02255}{{\ttfamily
  2105.02255}}].

\bibitem{Martinec:2022okx}
E.~J. Martinec, S.~Massai and D.~Turton, \emph{{On the BPS sector in
  AdS\_3/CFT\_2 Holography}},
  \href{https://arxiv.org/abs/2211.12476}{{\ttfamily 2211.12476}}.

\bibitem{Bufalini:2022wyp}
D.~Bufalini, S.~Iguri, N.~Kovensky and D.~Turton, \emph{{Worldsheet Correlators
  in Black Hole Microstates}},
  \href{https://doi.org/10.1103/PhysRevLett.129.121603}{\emph{Phys. Rev. Lett.}
  {\bfseries 129} (2022) 121603}
  [\href{https://arxiv.org/abs/2203.13828}{{\ttfamily 2203.13828}}].

\bibitem{Bufalini:2022wzu}
D.~Bufalini, S.~Iguri, N.~Kovensky and D.~Turton, \emph{{Worldsheet computation
  of heavy-light correlators}},
  \href{https://doi.org/10.1007/JHEP03(2023)066}{\emph{JHEP} {\bfseries 03}
  (2023) 066} [\href{https://arxiv.org/abs/2210.15313}{{\ttfamily
  2210.15313}}].

\bibitem{Kutasov:2001uf}
D.~Kutasov, \emph{{Introduction to little string theory}}, {\emph{ICTP Lect.
  Notes Ser.} {\bfseries 7} (2002) 165}.

\bibitem{Giveon:1999tq}
A.~Giveon and D.~Kutasov, \emph{{Comments on double scaled little string
  theory}}, \href{https://doi.org/10.1088/1126-6708/2000/01/023}{\emph{JHEP}
  {\bfseries 01} (2000) 023}
  [\href{https://arxiv.org/abs/hep-th/9911039}{{\ttfamily hep-th/9911039}}].

\bibitem{Giveon:2017myj}
A.~Giveon, N.~Itzhaki and D.~Kutasov, \emph{{A solvable irrelevant deformation
  of AdS$_{3}$/CFT$_{2}$}},
  \href{https://doi.org/10.1007/JHEP12(2017)155}{\emph{JHEP} {\bfseries 12}
  (2017) 155} [\href{https://arxiv.org/abs/1707.05800}{{\ttfamily
  1707.05800}}].

\bibitem{Asrat:2017tzd}
M.~Asrat, A.~Giveon, N.~Itzhaki and D.~Kutasov, \emph{{Holography Beyond AdS}},
  \href{https://doi.org/10.1016/j.nuclphysb.2018.05.005}{\emph{Nucl. Phys. B}
  {\bfseries 932} (2018) 241}
  [\href{https://arxiv.org/abs/1711.02690}{{\ttfamily 1711.02690}}].

\bibitem{Chakraborty:2018vja}
S.~Chakraborty, A.~Giveon and D.~Kutasov, \emph{{$ J\overline{T} $ deformed
  CFT$_{2}$ and string theory}},
  \href{https://doi.org/10.1007/JHEP10(2018)057}{\emph{JHEP} {\bfseries 10}
  (2018) 057} [\href{https://arxiv.org/abs/1806.09667}{{\ttfamily
  1806.09667}}].

\bibitem{Chakraborty:2019mdf}
S.~Chakraborty, A.~Giveon and D.~Kutasov, \emph{{$T\bar{T}$, $J\bar{T}$,
  $T\bar{J}$ and String Theory}},
  \href{https://doi.org/10.1088/1751-8121/ab3710}{\emph{J. Phys. A} {\bfseries
  52} (2019) 384003} [\href{https://arxiv.org/abs/1905.00051}{{\ttfamily
  1905.00051}}].

\bibitem{Apolo:2019zai}
L.~Apolo, S.~Detournay and W.~Song, \emph{{TsT, $T\bar{T}$ and black strings}},
  \href{https://doi.org/10.1007/JHEP06(2020)109}{\emph{JHEP} {\bfseries 06}
  (2020) 109} [\href{https://arxiv.org/abs/1911.12359}{{\ttfamily
  1911.12359}}].

\bibitem{Georgescu:2022iyx}
S.~Georgescu and M.~Guica, \emph{{Infinite $\mathrm{T\bar T}$-like symmetries
  of compactified LST}},  \href{https://arxiv.org/abs/2212.09768}{{\ttfamily
  2212.09768}}.

\bibitem{Zamolodchikov:1986bd}
A.~Zamolodchikov and V.~Fateev, \emph{{Operator Algebra and Correlation
  Functions in the Two-Dimensional Wess-Zumino SU(2) x SU(2) Chiral Model}},
  {\emph{Sov. J. Nucl. Phys.} {\bfseries 43} (1986) 657}.

\bibitem{Kutasov:1998zh}
D.~Kutasov, F.~Larsen and R.~G. Leigh, \emph{{String theory in magnetic
  monopole backgrounds}},
  \href{https://doi.org/10.1016/S0550-3213(99)00144-3}{\emph{Nucl. Phys. B}
  {\bfseries 550} (1999) 183}
  [\href{https://arxiv.org/abs/hep-th/9812027}{{\ttfamily hep-th/9812027}}].

\bibitem{Teschner:1997ft}
J.~Teschner, \emph{{On structure constants and fusion rules in the SL(2,C) /
  SU(2) WZNW model}},
  \href{https://doi.org/10.1016/S0550-3213(99)00072-3}{\emph{Nucl. Phys.}
  {\bfseries B546} (1999) 390}
  [\href{https://arxiv.org/abs/hep-th/9712256}{{\ttfamily hep-th/9712256}}].

\bibitem{Giveon:2001up}
A.~Giveon and D.~Kutasov, \emph{{Notes on AdS(3)}},
  \href{https://doi.org/10.1016/S0550-3213(01)00573-9}{\emph{Nucl. Phys.}
  {\bfseries B621} (2002) 303}
  [\href{https://arxiv.org/abs/hep-th/0106004}{{\ttfamily hep-th/0106004}}].

\bibitem{Seiberg:1999xz}
N.~Seiberg and E.~Witten, \emph{{The D1 / D5 system and singular CFT}},
  \href{https://doi.org/10.1088/1126-6708/1999/04/017}{\emph{JHEP} {\bfseries
  04} (1999) 017} [\href{https://arxiv.org/abs/hep-th/9903224}{{\ttfamily
  hep-th/9903224}}].

\bibitem{Eberhardt:2018vho}
L.~Eberhardt and K.~Ferreira, \emph{{Long strings and chiral primaries in the
  hybrid formalism}},
  \href{https://doi.org/10.1007/JHEP02(2019)098}{\emph{JHEP} {\bfseries 02}
  (2019) 098} [\href{https://arxiv.org/abs/1810.08621}{{\ttfamily
  1810.08621}}].

\bibitem{Dei:2021yom}
A.~Dei and L.~Eberhardt, \emph{{String correlators on AdS$_{3}$: four-point
  functions}}, \href{https://doi.org/10.1007/JHEP09(2021)209}{\emph{JHEP}
  {\bfseries 09} (2021) 209}
  [\href{https://arxiv.org/abs/2107.01481}{{\ttfamily 2107.01481}}].

\bibitem{Dei:2022pkr}
A.~Dei and L.~Eberhardt, \emph{{String correlators on $\text{AdS}_3$: Analytic
  structure and dual CFT}},
  \href{https://doi.org/10.21468/SciPostPhys.13.3.053}{\emph{SciPost Phys.}
  {\bfseries 13} (2022) 053}
  [\href{https://arxiv.org/abs/2203.13264}{{\ttfamily 2203.13264}}].

\bibitem{Berkovits:1999im}
N.~Berkovits, C.~Vafa and E.~Witten, \emph{{Conformal field theory of AdS
  background with Ramond-Ramond flux}},
  \href{https://doi.org/10.1088/1126-6708/1999/03/018}{\emph{JHEP} {\bfseries
  03} (1999) 018} [\href{https://arxiv.org/abs/hep-th/9902098}{{\ttfamily
  hep-th/9902098}}].

\bibitem{Cagnacci:2013ufa}
Y.~Cagnacci and S.~M. Iguri, \emph{{More $AdS_3$ correlators}},
  \href{https://doi.org/10.1103/PhysRevD.89.066006}{\emph{Phys. Rev. D}
  {\bfseries 89} (2014) 066006}
  [\href{https://arxiv.org/abs/1312.3353}{{\ttfamily 1312.3353}}].

\bibitem{Teschner:1999ug}
J.~Teschner, \emph{{Operator product expansion and factorization in the H+(3)
  WZNW model}},
  \href{https://doi.org/10.1016/S0550-3213(99)00785-3}{\emph{Nucl. Phys.}
  {\bfseries B571} (2000) 555}
  [\href{https://arxiv.org/abs/hep-th/9906215}{{\ttfamily hep-th/9906215}}].

\end{thebibliography}\endgroup

\end{document}